\documentclass[final,5p,times,twocolumn]{elsarticle}
\usepackage{graphicx}
\usepackage{subfigure}
\usepackage{epstopdf}
\usepackage{color}
\usepackage{multirow}
\usepackage{textcomp,gensymb}
\usepackage{lineno}
\usepackage{xcolor}
\usepackage{physics}
\usepackage{epsfig}
\usepackage{amsmath}
\usepackage{amssymb}
\usepackage{booktabs}
\usepackage{siunitx}
\usepackage{breqn}
\biboptions{square,comma,sort&compress}
\journal{}
\begin{document}
\begin{frontmatter}
\title{%A diffuse interface approach to study precipitate migration under temperature gradient
%Insight into migration of precipitate under thermal gradient - A phase-field approach
Interplay between thermal and compositional gradients decides the microstructure during thermomigration: a phase-field study  
}
%Understanding Thermomigration through a Novel Phase-Field Model: From Single-Phase Evolution to Multi-Phase Dynamics in Solid-State Alloys

\author[a,b]{Sandip Guin$^\dagger$}
\author[a,c]{Soumya Bandyopadhyay$^\dagger$}
\author[d]{Saswata Bhattacharyya\corref{cor1}}
\ead{saswata@msme.iith.ac.in}
\author[a]{Rajdip Mukherjee\corref{cor1}}
\ead{rajdipm@iitk.ac.in}
\cortext[cor1]{Corresponding Authors}

\address[a]{Department of Materials Science and Engineering, Indian Institute of
Technology, Kanpur, Kanpur-208016, UP, India}

\address[b]{International College of Semiconductor Technology, National Yang Ming Chiao Tung University, Hsinchu 300, Taiwan} 

\address[c]{Department of Materials Science and Engineering, University of Florida, Gainesville, Florida-32611} 

%\address[c]{Department of Materials Science and Engineering, National Yang Ming Chiao Tung University, Hsinchu 300, Taiwan}
\address[d]{Department of Materials Science And Metallurgical Engineering
, Indian Institute of Technology, Hyderabad
, Sangareddy - 502285, Telangana, India}
%\authorrunning{Short form of author list} % if too long for running head

\begin{abstract}
The presence of thermal gradients in alloys often leads to non-uniformity in concentration profiles, which can induce the thermomigration of microstructural features such as precipitates. 
To investigate such microstructural changes, we present a phase-field model that incorporates coupling between concentration and thermal gradients. First, we simulated the evolution 
of non-uniform concentration profiles in the single-phase regions of Fe-C and Fe-N alloy systems due to imposed thermal gradients.  
To validate our model with the classical experiments performed by Darken and Oriani,
we studied the evolution of spatially varying concentration profiles where 
thermal gradients encompass single-phase and two-phase regions. 
We developed a parameterized thermodynamic description of the two-phase region of a binary alloy to systematically study the effect of interactions between chemically-driven and thermal gradient-driven diffusion of solute on the evolution of precipitates. 
Our simulations show how thermal gradient, precipitate size, and interparticle distance influence the migration and associated morphological changes of precipitates. The composition profiles and migration rates obtained from single-particle simulations show an exact match with our analytical model. 
We use two-particle simulations to show conditions under which thermomigration induces the growth of the smaller particle and shrinkage of the larger one in contrast to the isothermal Ostwald ripening behavior. Our multiparticle simulations show similar behavior during coarsening. Moreover, in the presence of a thermal gradient, there is a shift in the center of mass of the precipitates towards the high-temperature region. \textcolor{black}{Thus, our study offers new insights into the phenomena of microstructure evolution in the presence of thermal gradient.}

\end{abstract}
\begin{keyword}
Phase-field Model \sep Thermomigration \sep Coarsening  \sep Microstructure
\end{keyword}
\end{frontmatter}
\def\thefootnote{$\dagger$}\footnotetext{These authors contributed equally to this work}\def\thefootnote{\arabic{footnote}}
\section{Introduction}
\label{intro}
Thermomigration, also known as the Soret effect or the Ludwig-Soret effect, is the phenomenon wherein a thermal gradient triggers mass transport~\cite{DARKEN1954841, asaro2008soret,costeseque2011thermodiffusion,steiner19571}. 
%This concept has captivated the scientific community for decades, piquing long-standing interest.
Ludwig first 
observed the evolution of concentration gradient in a liquid mixture due to thermal 
gradient~\cite{Ludwing_main}. Later, Soret reported a similar phenomenon in a solution of sodium chloride and potassium nitrate~\cite{Soret_main}. 
%Enhancing the quantitative comprehension of this migration process would offer advantages across various scientific and engineering fields. 
Thermomigration naturally observed in floating sea-water ice where a temperature gradient between seawater and solid ice leads to the removal of salt from the ice~\cite{Whitman1926}.

Though initially observed in liquid system, thermomigration is equally important in solid-state microstructural evolution.
%For example, thermomigration in lead-free solder joints in modern 3D integrated circuits can rapidly deteriorate device performance~\cite{xyz}. With the rapid development of miniaturized integrated devices with high packing density, the interconnects are exposed to larger thermal gradients, leading to severe reliability issues.   
Thermomigration causes the translation migration of microstructural features such as precipitates, voids and inclusions toward hotter or colder ends subject to a positive entropy production rate.
For instance, in nuclear fuel cells, nanosized gas bubbles migrate towards the higher temperature side due to large thermal gradients in the fuel cell~\cite{WANG2020109817, SENS1972293}. The chemical potential gradient can be related to the thermal gradient according to
$\mathbf{\nabla}\mu_{T} = -S^{T} c \frac{\mathbf{\nabla}T}{T}$,
where $S^{T}$ is the thermotransport coefficient and $c$ denotes the overall concentration of the migrating species~\cite{chen2019phase}.

\begin{figure*}[ht]
\centering 
\includegraphics[width=0.99\linewidth]{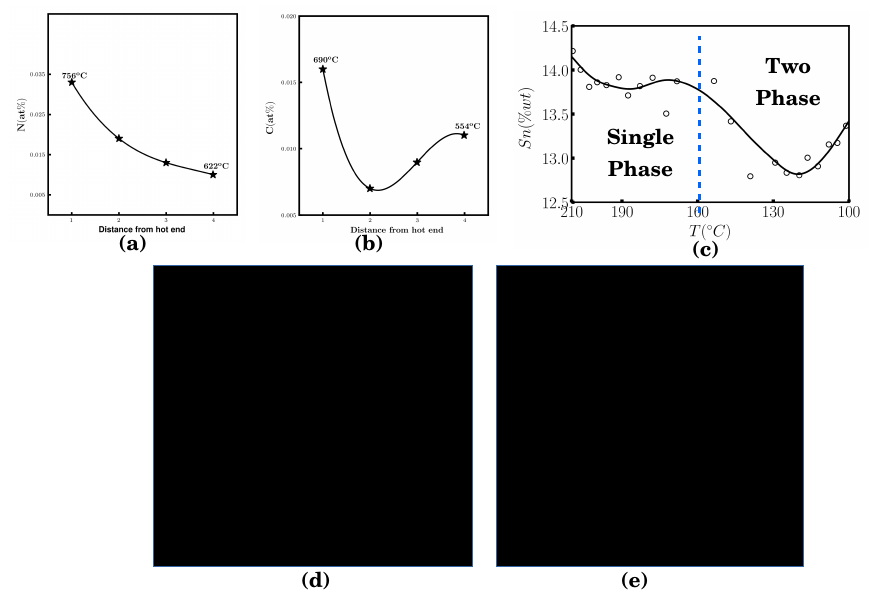}
\label{fig:intro}
\caption{(a) The nitrogen (N) concentration gradient in $\alpha$-Fe evolves due to the presence of a thermal gradient, as documented by Darken and Orani~\cite{DARKEN1954841}. (b) Similarly, the carbon (C) concentration gradient in $\alpha$-Fe system under the influence of a thermal gradient, as reported by Darken and Orani~\cite{DARKEN1954841}. 
(c) Sn concentration gradient formation in Pb-Sn alloys under thermal gradient as documented by Steiner ~\cite{steiner19571}. The higher temperature end is in the single-phase region, and the lower temperature end is in the two-phase region~\cite{steiner19571}. The blue-dotted vertical line separates the single-phase and two-phase regions. (d) Pt-Si droplets migrate on a Si (001) surface, where the periphery of the Si(001) surface has a lower temperature compared to the centre point. Figure will be displayed after publication with permission ~\cite{wcyang}. (e) Migration of Te precipitates in CdZnTe crystal. Figure will be displayed after publication with permission ~\cite{MEIER20094247}. The small red circle represents the Te precipitate. Initially located within the region bounded by two solid black lines (shown in the top image), some precipitates have migrated outside this region due to thermal gradient, as indicated in the bottom image. \textcolor{black}{In this case, the left interface of the precipitate matrix has a higher temperature, while the right interface has a lower temperature.}
}     % Give a unique label
\end{figure*}

Thermal gradient leads to the development of concentration gradient within a solid single-phase system, where solute migrates from lower to higher temperatures.
In an experimental monograph, Darken and Orani demonstrated evolution of composition gradient in the
presence of a thermal gradient in single-phase $\textrm{Fe-C}$ and $\textrm{Fe-N}$ alloy as shown in Figure~\ref{fig:intro}a and 
Figure~\ref{fig:intro}b, respectively~\cite{DARKEN1954841}. 
A temperature range of $622^oC$ to $756^oC$ leads to the evolution of the nitrogen (N) concentration gradient in an $\alpha$-Fe phase as shown by the solid black line in Figure~\ref{fig:intro}a. 
The black solid line in Figure~\ref{fig:intro}b illustrates the evolution of the carbon (C) concentration gradient in the $\alpha$-Fe phase over a 
temperature range spanning $554^oC$ to $690^oC$. In both alloys, the initial solute concentration was homogeneous. However, due to the 
thermal gradient in the material, solute atoms (C $\&$ N) migrate toward higher temperature regions, manifesting a gradient in the solute 
concentration~\cite{DARKEN1954841}. 

In a system transitioning from a single-phase to two-phase state, alterations in microstructural development occur due to thermal gradients. In this context, a single-phase to two-phase system denotes a material subjected to a thermal gradient where the high-temperature end resides within the single-phase region, while the low-temperature zone exists in the two-phase region.
In a classic experiment, Steiner demonstrated the evolution of the concentration gradient for the Pb-Sn system, where he considered the Pb-Sn alloys with an initial Sn  concentration of 14 wt\%~\cite{steiner19571}. In this study, the temperature range was $101^{\circ}C$ to $210^{\circ}C$. 
 As per the Pb-Sn phase diagram at 14 wt\% Sn, above $160^{\circ}C$, the alloy is in the single-phase region, and below that, the alloy is in the two-phase region~\cite{steiner19571}. 
Figure~\ref{fig:intro}c shows the final Sn concentration under thermal gradient for Pb-Sn alloy as per the experiment done by Steiner ( The blue dotted vertical  line separates the single phase from the two-phase region)~\cite{steiner19571}.

In the two-phase system, the second-phase particles exhibit temperature-dependent migration when subjected to a thermal gradient~\cite{Whitman1926}.
%from the 
%$eutectic temperature to the melting %%%temperature~\cite{Whitman1926}. 
For example,
Zappettini \textit{et. al.} used laser-induced thermal gradient to remove Te inclusion in CdZnTe crystal ~\cite{Zappettini}.
%In a two-phase alloy system where the equilibrium concentration of at \textcolor{green}{this line needs modification}least one phase depends on temperature, a thermal gradient prompts the migration of that particular phase toward the region of higher temperature~\cite{Whitman1926}.
Another work by Pawar \textit{et al.} investigated the migration of liquid particles within a solid phase towards the higher temperature, with the equilibrium concentration of the liquid phase being temperature-dependent, while the equilibrium concentration of the solid phase remains constant regardless of the temperature change~\cite{PAWAR2022117780}.
Wang \textit{et al.} presented an observation of Pt-Si droplets migrating on a Si $(001)$ substrate driven by a thermal 
gradient~\cite{wcyang}.
As shown in Figure~\ref{fig:intro}d, the Pt-Si droplets migrate towards the center of the Si substrate 
during annealing due to the presence of a thermal gradient between the substrate centre (higher temperature region) and its edge 
(lower temperature region). Consequently, this thermal gradient generates a Si concentration gradient within the Pt-Si droplets, with 
higher Si concentration at the higher temperature side. %This dictates the diffusion of Si from a higher-temperature interface to a lower one. 
This dictates the droplet migration is due to the ``dissolution-diffusion-deposition" flow of Si through the droplets~\cite{wcyang}. 
The experimental observation of Meier \textit{et al.} as shown in Figure~\ref{fig:intro}e, indicates the migration 
of the Te precipitates (marked as red circles within the two black lines) in CdZnTe 
crystal due to the presence of thermal gradient inside the precipitates~\cite{MEIER20094247}.
In Figure~\ref{fig:intro}e, the top image represents the microstructure before applying the thermal gradient, with Te precipitates represented by red circles within the region shown by two black vertical lines.
\textcolor{black}{The bottom image represents the microstructure after applying the thermal gradient. In this case, a thermal gradient is applied using a laser source in such a way that the left interface of the precipitates has a higher temperature while the right interface of the precipitate has a lower temperature. 
 Here it is clearly visible that some precipitates have migrated beyond the region bounded by the two black vertical lines~\cite{MEIER20094247}.}

%In Figs.~\ref{fig:intro}(a-b), the black lines represent the experimental results obtained by Darken \textit{et. al.}, and the red lines correspond to our simulation results. The detailed analysis of this study will be discussed in the section~\ref{simdetails} \textcolor{red}{Please add this part in the simulation section.}. 

 While the experimental evidence for thermomigration exists in solid-state alloys, there is a scarcity of comprehensive theoretical frameworks~\cite{DARKEN1954841, MEIER20094247, WANG2020109817,steiner19571}. 
 The scientific community must study it adequately after the work done by Darken and Orani~\cite{DARKEN1954841}. 
Most studies to date mainly focus on the heat transport coefficient and are limited to void/defects 
migration; minimal investigation focuses on the possibility of thermomigration arising from equilibrium composition 
gradient in the precipitate-matrix system~\cite{chen2019phase}. 
Moreover, no significant approach 
focuses on the fundamental investigation of the origin of composition gradient due to the presence of thermal gradient as experimentally validated by Darken and Orani.~\cite{DARKEN1954841}. 
Furthermore, in two-phase alloys, where temperature variation influences the equilibrium concentrations of both phases (interchange of the equilibrium concentration between solute-rich and solute-poor phases at higher temperatures), frames some open questions, such as 
how phase migration will manifest in a two-phase alloy characterized by temperature-dependent solubility for both phases under a thermal gradient? Another critical consideration is how this phase migration will impact the process of phase coarsening.

%In Figs.~\ref{fig:intro}(a-b), the black lines represent the experimental results obtained by Darken \textit{et. al.}, and the red lines correspond to our simulation results. The detailed analysis of this study will be discussed in the section~\ref{simdetails}.

%A broad spectrum of two-phase alloys can be found in which temperature variations influence the equilibrium concentrations of both phases. Take, for instance, the scenario within the two-phase region below the eutectic temperature in a eutectic system, where both the solvus lines exhibit temperature dependence. In such systems, the equilibrium concentration of the phase rich in solute diminishes as temperature rises, whereas the equilibrium concentration of the phase depleted in solute increases with elevated temperatures.
%However, no proper report exists 
%research has delved into 
%on the thermomigration within such two-phase alloys. This leaves one open question: How phase migration will manifest in a two-phase alloy characterized by temperature-dependent solubility for both phases under a thermal gradient? Additionally, another critical consideration is how this phase migration will impact the process of phase coarsening.  

 We introduce a phase-field model to answer all these questions. 
 Thus, this study aims to understand the complex synergism of compositional and 
 thermal interactions in different systems.
Phase-field modelling serves as a sophisticated computational tool for 
microstructure simulation, eliminating the need for explicit interface 
tracking~\cite{doi:10.1146/annurev.matsci.32.112001.132041}. Its continuous 
representation of field variables, coupled with a diffuse interface region, 
facilitates precise simulations of complex microstructural 
evolution~\cite{VERMA2023119393,Bandyopadhyay_2023,FEYEN2023119087}. For example, 
using phase-field modelling, Chakraborty \emph{et al.} studied grain boundary 
grooving in the presence of both electric current and thermal 
gradient~\cite{CHAKRABORTY2018377}. Attarti \emph{et al.} studied the thermal 
gradient-driven bonding process in 3D IC solder joint using phase-field 
simulation ~\cite{9091578}.

 Our analysis starts with studying the thermomigration in a single-phase binary system, following Darken and Orani~\cite{DARKEN1954841}.
We then explore a situation in which we witness a shift from a single-phase to two-phase state, with the higher temperature region constituting the single-phase region and the lower temperature region representing the two-phase region~\cite{steiner19571}.
The simulations related to single-phase and single-phase to two-phase transitions are mainly for validating our simulation results with existing experimental findings.
Finally, we study the thermomigration effect in two phase system.
We choose  precipitate-matrix system for the two-phase system to explore the precipitate migration mechanism through a matrix under a thermal gradient. We also explored the coarsening kinetics of precipitate under a thermal gradient. 
The following sections demonstrate the implemented model and detailed computational methodology.
We report our findings, analysis, and observations in the subsequent sections.

\section{Model formulation}
\subsection{Phase-field formulation}
\label{phase field model}
 %In this work, we introduce a novel phase-field model to study the thermal gradient-driven thermomigration in two-phase alloys where the equilibrium concentration of both phases depends on temperature. 
Phase-field modelling presents a simulation technique that enables a straightforward study of microstructural evolution under various conditions~\cite{KADIRVEL2023118438,VARMAR2023118529,WU2022117978,Bandyopadhyay_2023}. In this approach, we consider the interface to have a definite width, and the phase-field 
parameters change continuously across the width (See the comprehensive reviews by Chen~\cite{chen2002phase}, 
Steinbach~\cite{steinbach2009phase}, etc. for more details).
We define the free energy of a system in terms of the phase-field variables, which are subsequently relaxed to equilibrium~\cite{miralverma}.
Moreover, the temporal evolution of the order parameters ensures the equilibrium, hence mapping the microstructure or morphological changes in different phases~\cite{chen2002phase}.

The phase-field model has been used to study various microstructure evolution under external thermal, electrical, and mechanical fields. Using the phase-field method,
Mohanty \emph{et al.} studied the diffusion due to thermal gradient~\cite{mohanty2009diffusion}. Zhang \emph{et al.} investigated the pore migration 
under thermal gradient coupled with heat transfer~\cite{zhang2012phase}. Recently, Attari \emph{et al.}, using ``CALPHAD-reinforced multiphase-field model",
investigated the thermal gradient
assisted directional solidification in Cu-Sn micro solder units for rapid bonding in 3-D IC packaging~\cite{attari2020phase}.
The effect of interfacial anisotropy on voids/defects migration 
has also been extensively studied in the last decade~\cite{chen2019phase}. 
Liang \emph{et al.} investigated the effect of anisotropy in thermal conductivity on the migration of grain
boundaries~\cite{liang2021effect}.
So, phase-field modelling is an efficient tool for studying microstructure evolution under thermal gradient. 
Thus, we also use a phase-field model to study the effect of thermal gradient on microstructure evolution. 

Let us assume a model system where a precipitate and matrix phase coexist at equilibrium.
In this study, we can describe the complete microstructure of the system considering a spatial and temporal 
distribution of concentration field $c(\mathbf{r},t)$. 
Hence, we choose concentration $c(\mathbf{r},t)$ to be the phase-field variable (conserved), where $\mathbf{r} 
= (x, y, z)$ denotes the spatial coordinates 
in the Cartesian frame of reference, and $t$ denotes time. Note that here, we use bold letters to denote the 
vector fields. 

To understand the effect of thermal gradient on mass transport, let us start with the force-flux relationship based on irreversible thermodynamics in a binary system. As already
mentioned in this study we have used concentration $c(\mathbf{r},t)$ as the conserved order parameter; we adopt the Cahn-Hilliard approach in this model~\cite{cahn1961spinodal}.

\begin{figure}[htpb]
\centering 
   \includegraphics[width=1.0\linewidth]{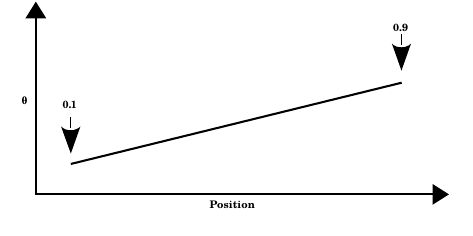}
\caption{Scaled temperature ($\theta$) profile throughout the domain. Right hand side has high temperature ($\theta_{high}$=0.9) and left hand side has low temperature ($\theta_{low}$=0.1).}
\label{fig:T_plot}      % Give a unique label
\end{figure}

\begin{figure}[htpb]
\centering 
\subfigure[${\Delta}f_{mix}(c(\mathbf{r},t), \theta)$ surface plot]
{\includegraphics[width=1.0\linewidth]{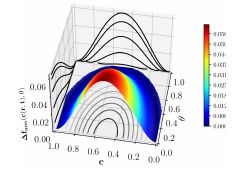}
\label{fig:model_energy_surface}}
\subfigure[${\Delta}f_{mix}(c(\mathbf{r},t), \theta)$ vs c]
{\includegraphics[width=0.8\linewidth]{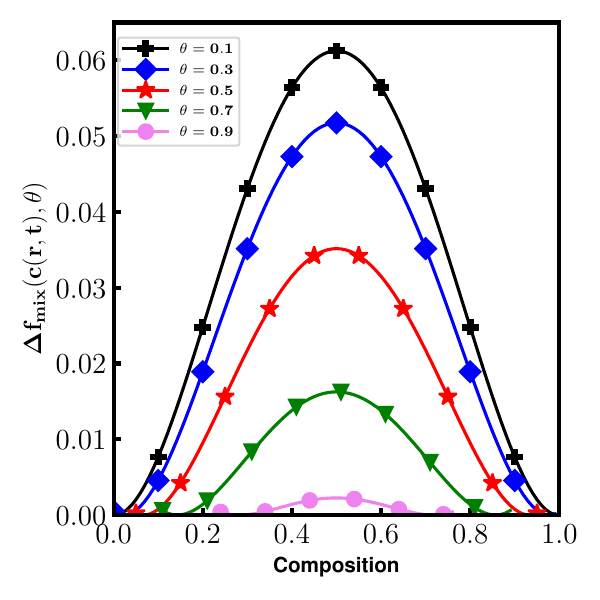}
\label{fig:model_f_c}
}
\subfigure[phase diagram ]
{\includegraphics[width=0.8\linewidth]{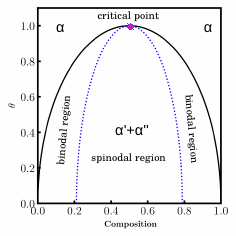}
\label{fig:model_phase_diagram}
}
\caption{(a) Free energy (${\Delta}f_{mix}(c(\mathbf{r},t), \theta)$) surface plot as a function of composition(c) and temperature. (b) (${\Delta}f_{mix}(c(\mathbf{r},t), \theta)$) vs composition (c) at different temperature ($\theta$). (c) Phase diagram derived from (${\Delta}f_{mix}(c(\mathbf{r},t), \theta)$).  
}
% figure caption is below the figure
\label{Fig:2}      % Give a unique label
\end{figure}

   For an isothermal inhomogeneous system, if only the internal process is the redistribution of the chemical species $i$, mass conservation requires:
\begin{equation}
   \frac{\partial c_i(\mathbf{r},t)}{\partial t} = -\mathbf{\nabla}\cdot \mathbf{J}_i,
\label{eqn:1}
\end{equation}
where $\mathbf{J}_i$ is the chemical transport flux density. Assuming linear kinetics we have:
\begin{equation}
   \mathbf{J}_i = -M_{ij}c_{0}\mathbf{\nabla} \mu_{j}.
\label{eqn:2}
\end{equation}
Here $M_{ij}$ is the mobility matrix describing the magnitude of the flux density of species $i$ due to the chemical potential gradient of species $j$, and $c_0$ defines the average concentration of the system.
Combining Equation~\eqref{eqn:1} and Equation~\eqref{eqn:2} we obtain,
\begin{equation}
  \frac{\partial c_i(\mathbf{r},t)}{\partial t} = -\mathbf{\nabla}\cdot \left[-M_{ij}\mathbf{\nabla} \mu_{j}\right].
\label{eqn:3}
\end{equation}

In the phase-field model, we describe the quantification of chemical energy dissipation 
through the evolution of compositional and internal interfacial areas, which is expressed in relation to the 
overall free energy functional~\cite{cahn1961spinodal}.
Thus, we consider the dependency of the local free energy to be not only on the local composition $c_i$ but also on the gradient
$\mathbf{\nabla}c_i(\mathbf{r},t)$ as
\begin{subequations}
\begin{equation}
    F_{total} = F_{bulk} + F_{gradient},
    \label{eqn:4a}
    \end{equation}
    \begin{equation}
          F_{total} = \int_{V}[f(c_{i}(\mathbf{r},t)) + \frac{\kappa_{c_i}}{2}|\nabla c_{i}|^2 ]dV.
          \label{eqn:4b}
    \end{equation}
\end{subequations}

Here, we define the total free energy $F_{total}$ as the sum of bulk free energy $(F_{bulk})$ and the gradient free energy $(F_{gradient})$.

For simplicity, we approximate the 
gradient energy up to the second order. We describe $f(c_i(\mathbf{r},t))$ as the bulk free energy density, and $\kappa_{c_i}$ as the
gradient energy coefficient (can be expanded to higher orders to incorporate anisotropy related to crystallographic symmetry)
associated with the system.

  The chemical potential relative to the final equilibrium state is defined as
 \begin{equation}
   \begin{split}
      \mu_{i} &= \frac{\delta F_{total}}{\delta c_{i}} \\
             &= \frac{\partial f(c_{i}(\mathbf{r},t))}{\partial (c_{i}(\mathbf{r},t))} - \kappa_{c_i}\nabla^{2}c_{i}(\mathbf{r},t).
         \end{split}
\label{eqn:5}
 \end{equation}
 
Putting the expression of $\mu_{i}$ (Equation\eqref{eqn:5}) into Equation~\eqref{eqn:3} we obtain:
\begin{equation}
\begin{split}
  \frac{\partial c_i(\mathbf{r},t)}{\partial t} &= -\mathbf{\nabla}\cdot \left[-M_{ij}\mathbf{\nabla}\frac{\delta F_{total}}{\delta c_{i}}\right] \\
                                                &= \mathbf{\nabla}\cdot M_{ij}\left[\mathbf{\nabla}\left(\frac{\partial f(c_{j}(\mathbf{r},t))}{\partial (c_{j}(\mathbf{r},t))}
                                                - \kappa_{c_j}\nabla^{2}c_{j}(\mathbf{r},t)\right)\right].
\end{split}
\label{eqn:6}
\end{equation}

In the presence of thermal gradient when mass flow is dictated by the thermo-transport mechanism we rewrite Equation\eqref{eqn:1} as
\begin{equation}
   \frac{\partial c_i(\mathbf{r},t)}{\partial t} = -\mathbf{\nabla}\cdot \mathbf{J}_i -\mathbf{\nabla}\cdot \mathbf{J}^{i}_{Q}.
\label{eqn:7}
\end{equation}

Here $\mathbf{J}_{Q}^{i}$ defines the thermal mass flow density or the thermal flux of the chemical species $i$. We rewrite $\mathbf{J}_{Q}^{i}$
as:
\begin{equation}
 \mathbf{J}^{i}_{Q} = -\frac{M_{ij}^{T}Q c_j(\mathbf{r},t)\mathbf{\nabla} T}{T},
\label{eqn:8}
\end{equation}
where $M_{ij}^{T}$ is the thermal mobility, $T$ denotes the temperature, and $Q$ is the thermo-transport heat often considered as the change of heat, related to the solute particle transfer between two regions at slightly different temperatures~\cite{eastman1926thermodynamics,zhao2021ionic}. Considering the unit of $Q$ to be \textrm{$\frac{J}{mole}$}~\cite{chen2012thermomigration}, $M_{ij}^{T}$ as $\frac{m^2 mole}{Js}$, and $c_i$ to be $\frac{mole}{m^3}$, Equation~\ref{eqn:8} becomes dimensionally consistent with thermal flux $\mathbf{J}^{i}_{Q}$ having unit of $\frac{mole}{m^2 s}$ (similar to $\mathbf{J}^{i}$ in Equation~\ref{eqn:2}). 
Thus, the total flux $\mathbf{J}$ can be rewritten as:
%\begin{equation}
%    \begin{split}
%    \mathbf{J_i} &= \mathbf{J}_i + \mathbf{J}^{i}_{Q}\\
%               &= -D_{ij}\mathbf{\nabla}c_j - D^{T}_{ij}c_j\mathbf{\nabla}T, 
%    \end{split}
%    \label{eqn:9}
%\end{equation}
\begin{equation}
    \begin{split}
    \mathbf{J} &= \mathbf{J}_i + \mathbf{J}^{i}_{Q}\\
               &= -M_{ij}\mathbf{\nabla} \mu_{j} -\frac{M_{ij}^{T}Q c_j(\mathbf{r},t)\mathbf{\nabla} T}{T}. 
    \end{split}
    \label{eqn:9}
\end{equation}

When mass flow vanishes ($\mathbf{J} = 0$) the chemical potential gradient is given 
as:
\begin{equation}
 \mathbf{\nabla}\mu_{i} = -S^{T}_{ij}c_{j}\frac{\mathbf{\nabla}T}{T},
 \label{eqn:10}
\end{equation}
here, $S^{T}_{ij} = \frac{M_{ij}^{T}}{M_{ij}}Q$ is the thermo-transport coefficient and its sign indicates the direction of the flux of the solute. For example positive $S^{T}_{ij}$ means the solute migration occurs toward the
colder terminal~\cite{mohanty2009diffusion}.

Together Equation~\eqref{eqn:7} and Equation~\eqref{eqn:10} provide the equation for mass transport due to the presence of thermal gradient as:

\begin{equation}
\begin{split}
  \frac{\partial c_i(\mathbf{r},t)}{\partial t} &= \mathbf{\nabla}\cdot M_{ij}\left\{\left[\mathbf{\nabla}\left(\frac{\partial f(c_{j}(\mathbf{r},t))}{\partial (c_{j}(\mathbf{r},t))}
                                                - \kappa_{c_j}\nabla^{2}c_{j}(\mathbf{r},t)\right)\right]
                                                + \left[S^{T}_{ij}c_{j}\frac{\mathbf{\nabla}T}{T}\right]\right\}.
\end{split}
\label{eqn:11}
\end{equation}

%\{However,
%in this study, we assume a system \textcolor{green}{rewrite this part, something is missing}in a certain temperature region where $S^{T}_{ij} \approx 0.0$ 
%and formulate a new toy model, where we consider the equilibrium compositions of the chemical species as a function of temperature $T$.\}

%Moreover, we also comment that since this is a simplistic model, assuming composition as a function of the temperature
%gives rise to an equivalent scenario of using the thermo-transport equation. Although, in a real situation, we need to consider the thermo-transport 
%(Equation~\eqref{eqn:8}) as a separate contribution, since discerning the exact coupling of the equilibrium composition on the temperature is intricate
%and convoluted.

We have discussed earlier that temperature-dependent equilibrium solubility also causes the thermomigration of distinct phases. 
The primary objective of this paper is to investigate this thermomigration driven by temperature-dependent 
equilibrium solubility. Thus, we focus on a temperature range where the system is diffusion-dominated (considering higher mass flow)
and the effect arising due to the thermal flux $\mathbf{J}^{i}_{Q}$ is negligible or assuming 
$M^{T}_{ij} \approx 0.0$ in Equation~\eqref{eqn:10}.
Moreover, we introduce a new model that considers a two-phase solid solution within this 
temperature range, with the equilibrium solute concentration of each phase as a function of temperature T.

In our simulation we use scaled form of all the variable. For example, we use scaled form of temperature. This scaled temperature is represented by $\theta$. We also assume a linear temperature distribution as shown in Figure~\ref{fig:T_plot}. 
We can obtain this 
the linear distribution of the temperature by solving steady state heat equation $\mathbf{\nabla}\cdot(k_{Q}\mathbf{\nabla} \theta) = 0$, where $k_Q$ denotes the thermal conductivity.
In general, $k_Q$ can be composition dependent, but here we have assumed it to be a constant ($k_Q = 1$).

To implement the model, we consider a double well potential for the bulk free energy as:

\begin{equation}
  f(c_{i}(\mathbf{r},t), \theta) = {\Delta}f_{mix}(c_{i}(\mathbf{r},t), \theta) = A(c_i-c_{eq_i}^{m}(\theta))^2(c_i - c_{eq_i}^{p}(\theta))^2,
\label{eqn:12}
\end{equation}
where, ${\Delta}f_{mix}(c_{i}(\mathbf{r},t), \theta)$ represents the double well potential shown in 
Figure~\ref{fig:model_energy_surface}, and A is a constant describing the barrier height of the assumed double 
well potential. For our simulation, we take A to be 1.0.

Figure~\ref{fig:model_f_c} shows the ${\Delta}f_{mix}(c(\mathbf{r},t), \theta)$ vs composition ($c_{i}
(\mathbf{r},t)$) diagram for different temperature, 
where we observe a reduction in the barrier height as a function of temperature. 
Figure~\ref{fig:model_phase_diagram} shows the corresponding phase diagram of the model. We calculate the phase 
diagram by 
computing the solute solubility of both matrix and precipitate phases using the common tangent construction at 
every temperature.

Further, we formulate the equilibrium compositions for the matrix $c_{eq_i}^{m}(\theta)$ and precipitate $c_{eq_i}^{p}(\theta)$ phases as a function of
temperature ($\theta$) as:

\begin{align}
    &c_{eq_i}^{m}(\theta) = 0.5-\sqrt{\frac{1-\theta^2}{4}},\\
    &c_{eq_i}^{p}(\theta) = 0.5+\sqrt{\frac{1-\theta^2}{4}}.
    \label{eqn:1314}
\end{align}

Finally, the spatio-temporal evolution of the concentration field in the matrix as well as precipitate 
is governed by the conserved Cahn-Hilliard equation ansatz~\cite{chen2002phase,cahn1961spinodal}:
\begin{equation}
    \frac{\partial c_i(\mathbf{r},t)}{\partial t}= \mathbf{\nabla}\cdot M_{ij}(c_{j}(\mathbf{r},t), \theta)\left[\mathbf{\nabla}\left(\frac{\partial f(c_{i}(\mathbf{r},t), \theta)}{\partial (c_{j}(\mathbf{r},t))}
                                                - \kappa_{c_j}\nabla^{2}c_{j}(\mathbf{r},t)\right)\right],
\label{eqn:15}
\end{equation}
where $M_{ij}(c_{i}(\mathbf{r},t), \theta)$ is the mobility of the system which can be composition as well as temperature dependent. 

%which is also temperature as well as composition dependent.

\begin{figure}[ht]
\centering 
\includegraphics[width=1.0\linewidth]{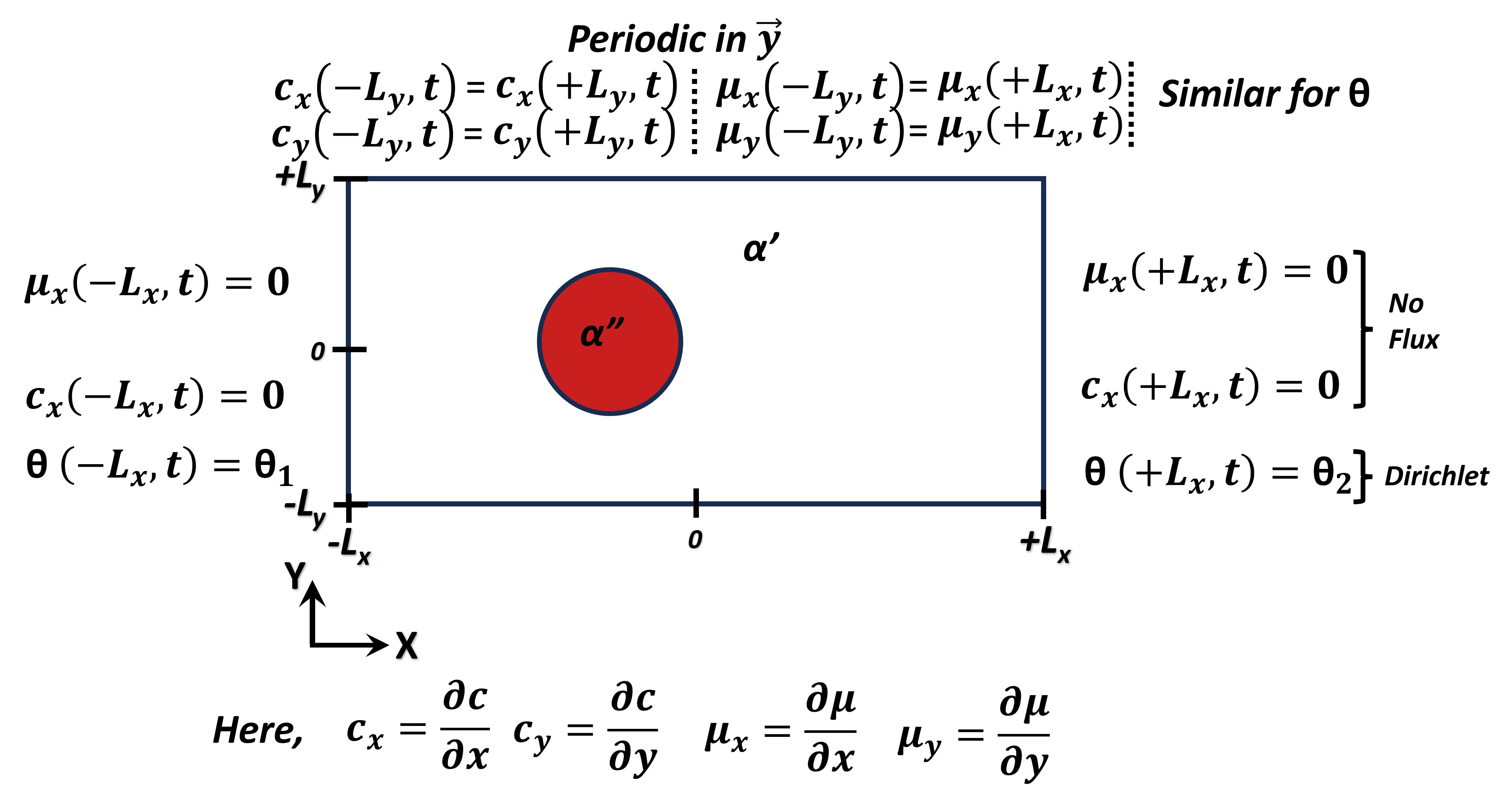}
\caption{Schematic shows domain boundary condition (BC). Here, $c$ denotes solute concentration, $\mu$ is the chemical potential and $\theta$ is scaled temperature. In our simulation $(\theta_2=\theta_{high}) > (\theta_1=\theta_{low})$. \textcolor{black}{In the Y-direction we employ periodic BC for composition (c), chemical potential($\mu$) and temperature ($\theta$). In the X-direction we employ no flux condition for both c and $\mu$, while for $\theta$, we employ Dirichlet BC.} }
\label{fig:bc_schematic}
\end{figure}

\subsection{Finite element discretization of the governing equations}

The finite element method (FEM) has been widely used to solve the complex partial differential equations (PDE)
with ease for many years by the scientific community due to simplicity in the discretization of the equations
into the weak form and efficiently handling the exact boundary conditions.
Here we perform a FEM based method to solve the 
governing phase-field equation (Equation~\eqref{eqn:15}) with proper boundary conditions. 
Moreover, in our case, the incorporation of the effect of temperature at the boundaries make this problem perfect 
to solve using FEM with mixed boundary conditions (Neumann, Dirichlet, and Periodic). 
\textcolor{black}{Figure~\ref{fig:bc_schematic} shows the details of boundary conditions (BC) for our simulation. In the Y-direction, we employ periodic BC for composition (c), chemical potential($\mu$) and temperature ($\theta$). In the X-direction, we employ no flux condition for both c and $\mu$, while for $\theta$, we employ Dirichlet BC. }
Although solving the Cahn-Hilliard equation is a bit difficult, 
due to the presence of gradient-free energy which is a fourth-order differential equation nevertheless, it can be solved in two ways.

First to prepare for the FEM discretization of Equation~\eqref{eqn:15}, 
we construct a weak form in a similar manner to that used by Stogner
et al.~\cite{stogner2008approximation}. The weighted integral residual projection of Equation~\eqref{eqn:15} 
is constructed using a test function $\phi_m$ and integrating
the second-order terms by parts once and the fourth order term
by parts twice. Thus after discretization Equation~\eqref{eqn:15} yields:
\begin{equation}
    \begin{split}
        \left(\frac{\partial c_i(\mathbf{r},t)}{\partial t}, \phi_{m}\right) &= -\left(\kappa_{c_{j}} \nabla^{2}c_j(\mathbf{r},t), \nabla\cdot(M_{ij}(c_{j}(\mathbf{r},t), \theta)\nabla\phi_{m})\right)\\
                                                           &-\left(M_{ij}(c_{j}(\mathbf{r},t), \theta) \nabla \left(\frac{\partial f(c_{j}(\mathbf{r},t)}{\partial c_{j}(\mathbf{r},t)}\right), \nabla \phi_{m}\right)\\
                                                           & +\langle M_{ij}(c_{j}(\mathbf{r},t), \theta) \nabla(\kappa_{c_{j}}(\nabla^2 c_{j}(\mathbf{r},t))\cdot \hat{n}), \phi_m \rangle\\
                                                           &- \left\langle M_{ij}(c_{j}(\mathbf{r},t), \theta)\nabla \left(\frac{\partial f(c_{j}(\mathbf{r},t)}{\partial c_{j}(\mathbf{r},t)}\right)\cdot \hat{n}, \phi_m \right \rangle\\
                                                           &+\langle \kappa_{c_{j}}\nabla^2 c_{j}(\mathbf{r},t), M_{ij}(c_{j}(\mathbf{r},t), \theta)\nabla\phi_m\cdot \hat{n} \rangle,
    \end{split}
    \label{eqn:16}
\end{equation}
where $(*,*)$ operator represents a volume integral with an inner product and $\langle *,* \rangle$ operator denotes the
surface integral with an inner product. 

  Another way to solve Equation~\eqref{eqn:15} is to split
the fourth order equation into two second order equations, such that two variables are solved, the concentration $c_i(\mathbf{r},t)$  
and the chemical potential $\mu_{i}$.
In this case, the two residual equations are: 
\begin{subequations}
\begin{equation}
\begin{split}
    R_{\mu_{i}} &= \left(\frac{\partial c_i(\mathbf{r},t)}{\partial t}, \phi_m\right) + \left( M_{ij}(c_{j}(\mathbf{r},t)\nabla\mu_j, \mathbf{\nabla} \phi_m \right) \\
          &- \langle  M_{ij}(c_{j}(\mathbf{r},t)\nabla\mu_{j}\cdot\hat{n},\phi_m \rangle. 
   \label{eqn:17a}
   \end{split}
\end{equation}
\begin{equation}
  \begin{split}
     R_{c_{i}} &= (\nabla c_i(\mathbf{r},t),\nabla(\kappa_{c_i} \phi_m)) - \langle \nabla c_i(\mathbf{r},t)\cdot\hat{n}, \kappa_{c_{i}}\phi_m\rangle \\
               &-\left(\left(\frac{\partial f(c_{i}(\mathbf{r},t)}{\partial c_i(\mathbf{r},t)} - \mu_{i}\right),\phi_m\right). 
\end{split}
\label{eqn:17b}
\end{equation}
\label{eqn:17}
\end{subequations}
In this work we mainly adopt the second or the split formalism of Equation~\eqref{eqn:17a} $\&$ Equation~\eqref{eqn:17b} 
as this change improves the solve convergence without any impact in the solution.

 Thus, we discretize each equation in its weak form in a typical FEM way. 
 All the required FEM objects and architecture are provided by
 MOOSE~\cite{tonks2012object, lindsay2022moose}. We use a QUAD4 type 
 mesh element to discretize the system. Lagrange shape functions are used for the
conserved field variable, chemical potential, and any coupled variables e.g. 
temperature. The transient system of PDEs is integrated over
time implicitly. The system (Equations.~\eqref{eqn:17a}-~\eqref{eqn:17b}) is solved using Newton's method~\cite{tonks2012object,lindsay2022moose} with appropriate
boundary conditions shown in Figure~\ref{fig:bc_schematic}.
Additionally, we assume isotropic and constant values for the mobility ($M_{ij}(c_{i}(\mathbf{r},t), \theta)=M_c$) and gradient energy coefficient ($\kappa_{c_i}= \kappa_c $).

\begin{figure}[ht]
\centering 
\includegraphics[width=0.7\linewidth]{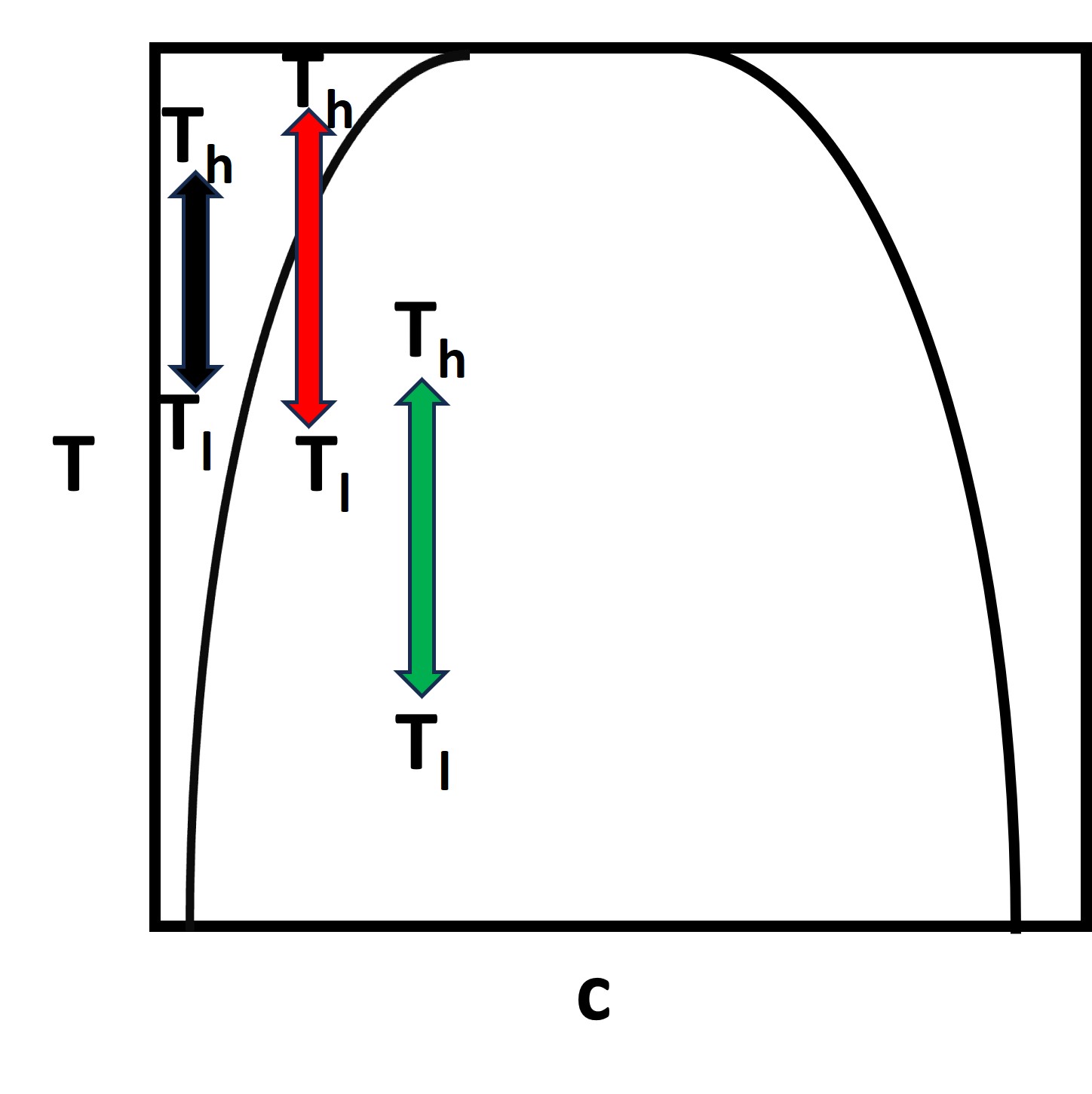}
\caption{ Schematic phase diagram showing three different sets of simulation. The black double-sided arrow shows the case where both high and low temperature is within the single phase region; the red double-sided arrow shows single phase to two-phase case, where high-temperature point is in the single-phase region and the low-temperature point is in the two-phase region. Finally the green double-sided arrow shows the two phase case, where both high and low temperature point is in the two phase region. }
\label{fig:phase_dia_1}
\end{figure}

\section{Model validation and simulation details}
\label{simdetails}

\begin{figure*}[ht]
\centering 
\includegraphics[width=0.7\linewidth]{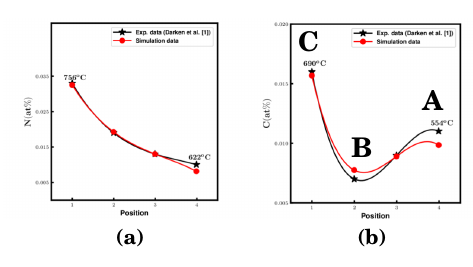}
\caption{(a) The nitrogen (N) concentration gradient in $\alpha$-Fe evolves due to the presence of a thermal gradient, as documented by Darken and Orani~\cite{DARKEN1954841}. (b) Similarly, the carbon (C) concentration gradient in $\alpha$-Fe undergoes changes under the influence of a thermal gradient, as reported by Darken and Orani~\cite{DARKEN1954841}. The experimental observations by Darken and Orani are represented by the black line, while our simulation results are depicted by the red line in both cases. In region-A, carbon diffusion comes to a halt due to the low temperature. In region-B, the carbon diffusivity is significant enough to allow solute migration. Consequently, carbon atoms move from region-B to region-A, where the temperature is higher.
}
\label{fig:darken_11}
\end{figure*}

\begin{figure}[ht]
\centering 
\includegraphics[width=0.8\linewidth]{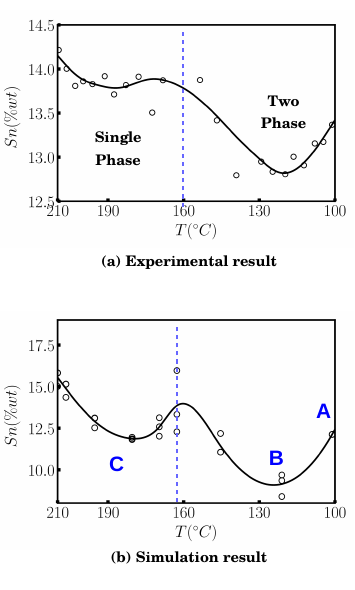}
\caption{(a) Shows the experimental observation (Sn concentration vs. temperature ($T^oC$)) done by Steiner for Pb-Sn system in single to two-phase region under a temperature range of $101^{\circ}C$ to $210^{\circ}C$ 
 ~\cite{steiner19571}. (b) Shows the our simulation results (Sn concentration vs. temperature ($T^oC$). }
\label{fig:single_two_pb_sn}
\end{figure}

%We use the MOOSE Framework to implement our model in the FEM environment. In the present study, we use a mixed 
%(Neumann - Periodic) boundary condition to solve Equations.~\eqref{eqn:17a}-~\eqref{eqn:17b} as shown in 
%Fig.~\ref{fig:bc_schematic}

%Additionally, we assume isotropic and constant values for the mobility ($M_{ij}$) and gradient energy coefficient ($\kappa_{c_i}$). Moreover, it is assumed that the thermal conductivity ($k_{th}$) of the matrix and the precipitate phases are the same and constant.

\subsection{Single-phase system}

In this sub section, we cross-reference our simulation outcomes against experimental data pertaining to single-phase system. In the context of a single-phase system, it involves a binary alloy experiencing a thermal gradient where both high and low-temperature points reside within the confines of the single-phase region. Essentially, the entire system remains within this single-phase domain, as visually indicated by the double-sided black arrow in Figure~\ref{fig:phase_dia_1}.
 To perform the simulations, we choose a single-phase 
$\alpha$-Fe in Fe-C and Fe-N systems following Darken and Orani~\cite{DARKEN1954841}. 
 %We perform these computations using the, which relies on Calphad principles
%We transform these free energy curves into a singular function considering both temperature and composition. We 
%utilize this resulting function to represent bulk free energy density ($f(c_i(\mathbf{r},t))$ in 
%equation~\ref{eqn:4b}), seamlessly integrating it into our phase-field simulations for the single-phase systems 
%of Fe-N and Fe-C.
More information regarding the simulations for is detailed 
in Section 1 of the Supplementary Material. 
%provides information regarding the free energy model and simulation specifics about Fe-N and Fe-C.

Figure~\ref{fig:intro}a shows the concentration gradient of N (at\%) within a temperature range $622^\circ C \to 
756^\circ C$. Figure~\ref{fig:intro}b shows the concentration gradient of C (at\%) within a temperature region 
$554^\circ C \to 690^\circ C$. 
The black and the red lines in both figures correspond to the experimental (done by Darken and Orani~\cite{DARKEN1954841}) and simulation results, 
respectively. As observed, our simulation results show a decent agreement between the concentration profiles of carbon (C) and nitrogen (N) within the $\alpha$-Fe phase and the experimental observations of Darken and Orani under thermal 
gradient~\cite{DARKEN1954841}.

%Figure~\ref{fig:intro}a shows the N (at\%) concentration gradient under temperature range $622^oC$ to 
%$756^oC$. 
%Figure~\ref{fig:intro}b shows the C (at\%) concentration gradient under temperature range $554^oC$ to $690^oC$. 
%The black and red lines in both figures correspond to the experimental and our simulation results, respectively.

In the regular solution model, the interplay between the increase in entropy and the enthalpy dictates the 
overall effect of increasing temperature on the thermodynamics (free energy) of mixing. Thus, 
for any system, changes in entropy 
$(\Delta S)$, enthalpy $(\Delta H)$, and hence free energy $(\Delta G)$ are mainly governed by the change in temperature.  
In general, an increase in temperature increases the entropy or the degree of disorder in a system, which can lead to a 
decrease in the free energy change of the system, as a higher entropy contribution term ($-T\Delta S$) can compensate for the 
enthalpy term~\cite{YIN2024119445}.   
This indicates the migration of solute atoms toward higher temperature regions (due to the lower free energy contribution at elevated temperatures), manifesting a concentration gradient in both the C and N inside the $\alpha$ ferrite phase. 
 Figure~\ref{fig:darken_11}a shows at higher temperatures, N concentration is higher, while at lower temperatures, N concentration is lower.

Figure~\ref{fig:darken_11}b shows the C concentration remains unchanged at the lower temperature regions (marked as region A in Figure~\ref{fig:darken_11}b). This is due to the diffusion of C atoms freezes due to lower diffusivity at the lower temperature region~\cite{DARKEN1954841}.
 %This is due the, at that temperature, diffusion of carbon atoms freezes due to significantly lower diffusivity.
 However, in the region where the temperature is sufficient for the diffusion of C atoms(marked as region B in Figure~\ref{fig:darken_11}b), the atoms start migrating toward the higher temperature region (marked as region C in Figure~\ref{fig:darken_11}b), which explains the dip in C concentration in the middle of the sample (marked as region B in Figure~\ref{fig:darken_11}b).

\subsection{Single-phase to two-phase system}

In this sub-section, we cross-reference our simulation outcomes against experimental data pertaining to single-phase to two-phase systems.
Here, the high-temperature point is situated within the single-phase region, while the low-temperature point falls within the two-phase region. This distinction is illustrated by the red arrow in Figure~\ref{fig:phase_dia_1}.

Steiner demonstrated the evolution of the concentration gradient for the Pb-Sn system, where they considered the Pb-Sn alloys with an initial Sn (wt\%) concentration of 14 wt\%~\cite{steiner19571}. In their study, the temperature range was $101^{\circ}C$ to $210^{\circ}C$. 
 As per the phase diagram  at 14 wt\% Sn, above $160^{\circ}C$, the alloy is in the single-phase region, and below that, the alloy is in the two-phase region~\cite{steiner19571}. 
Figure~\ref{fig:single_two_pb_sn}a shows the final Sn concentration after applying thermal gradient as done by Steiner for Pb-Sn alloy~\cite{steiner19571}. The blue vertical dotted line separates the single phase from the two-phase region.

Figure~\ref{fig:single_two_pb_sn}b shows the simulation results for the experiment performed by Steiner ~\cite{steiner19571}. 
We start our simulations with the initial Sn concentration of around 14 wt\%. 
Note that the initial microstructure of our simulations comprises both two-phase and single-phase regions. 
Within the single-phase region, the initial concentration of Sn is 14 wt\%.  Section 2 of the supplementary materials provides detailed simulation information.

With the rise in temperature within the two-phase regions, the Sn solubility of the matrix phase ($\alpha$) increases, while for precipitate ($\beta$) decreases. 
%It is important to note that the entire two-phase region is at equilibrium with the corresponding temperature.
\textcolor{black}{With time, due to the presence of thermal gradient, the solute within the single-phase region starts migrating toward regions with higher temperatures, resulting in Sn concentration gradient in the single-phase region, akin to the process observed in the case of Fe-N single-phase system as demonstrated by Darken and Orani~\cite{DARKEN1954841}. Thus, in ``Region C" of Figure~\ref{fig:single_two_pb_sn}b, we observe a decline in Sn concentration as the solute migrates toward higher temperature regions from this area. 
Consequently, the precipitate ($\beta$) in the two-phase region begins to dissolve, leading to the release of extra solutes. These solutes will subsequently migrate toward the single-phase region. Hence, the there is increase in Sn concentration near the single-phase and two-phase boundary region~\cite{steiner19571}.
As the precipitate dissolve in the two-phase region, the two-phase region become single phase region and in this region concentration gradient established. }

The region marked as ``Region A" in Figure \ref{fig:single_two_pb_sn}b (right-hand side of the plot) corresponds to the lowest temperature region. In this region,
the diffusion of solute atoms is arrested, resulting in an unchanged Sn concentration in the region. 
However, in ``Region B", the diffusivity is adequate to allow the diffusion of solute atoms from this region towards the higher temperature region. As a results there is a dip in Sn concentration in ``Region B"~\cite{steiner19571}.

\begin{figure}[ht]
\centering 
\includegraphics[width=0.99\linewidth]{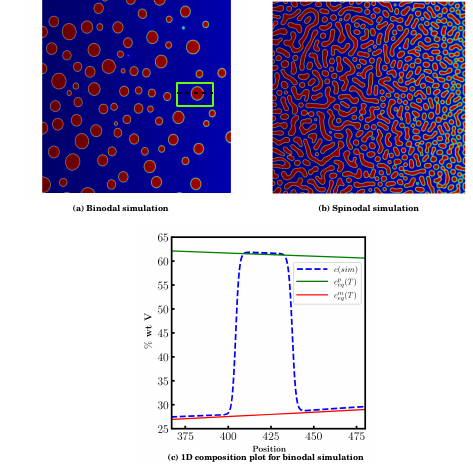}
\caption{(a) Shows the bimodal microstructure evolution for initial composition ($C_0$) of 29.43 wt $\%$ V. (b) Shows the spinodal microstructure evolution for initial composition ($C_0$) of 29.43 wt $\%$ V. (c) 1D composition profile taken along the black dotted line bounded by green rectangular box from the binodal microstructure along with comparison with the equilibrium concentration. \textcolor{black}{ Here, the blue line represents the 1D composition plot from the simulation, while the red and green lines represent the equilibrium V concentration in $\alpha$ and $\sigma$ phase, respectively.} }
\label{fig:darken_spino}
\end{figure}

\subsection{Two-phase system}

\begin{figure*}[ht]
\centering 
\includegraphics[width=1.0\linewidth]{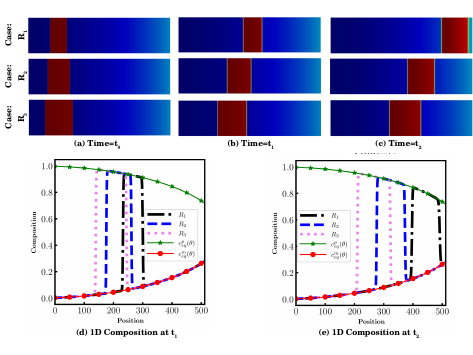}
\caption{Microstructure evolution of single flat interface precipitate simulation for all three cases ($R_1$ to $R_3$ : top to bottom) at (a) $time=t_o$ (initial time), (b) $time=t_1$, and (c) $time=t_2$. The red region is the precipitate, and the remaining blue region is the matrix phase. In all these simulations right end has the highest temperature, while the left end has the lowest temperature.
1D composition plot taken at $y=\frac{ny}{2}$ for all three cases along with analytical ($c_{eq}^{p}(\theta)$) and ($c_{eq}^{m}(\theta)$) at (c) $time=t_1$ and (d) $time=t_2$. Here, $t_o<t_1<t_2$.  }
\label{fig:R_S_1}% Give a unique label
\end{figure*}

\begin{figure}[htpb]
\centering 
\includegraphics[width=0.90\linewidth]{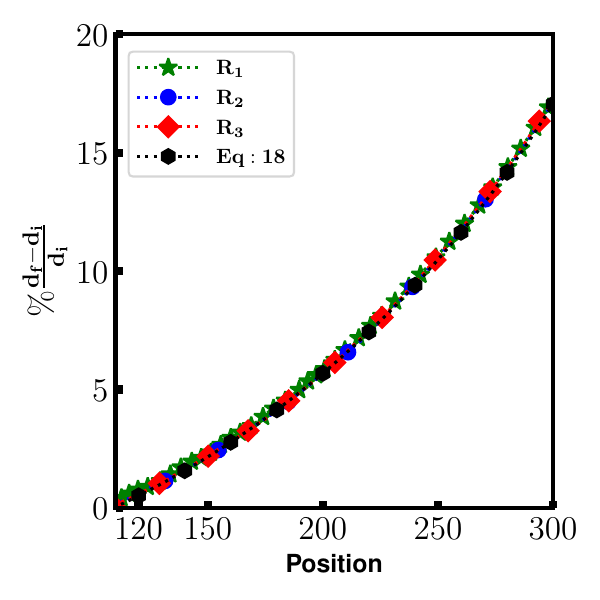}
\caption{ \% change in precipitate size wrt. initial size and comparrision with Equation~\ref{eq:size_chng} for all three flat interface single precipitate simulation cases ($R_1$, $R_2$ and $R_3$). \% change in precipitate size wrt. initial size for all three cases is the same and aligned with the theoretical calculations given by Equation~\ref{eq:size_chng}.  }
\label{fig:R_S_2}
\end{figure}

 We now simulate the microstructure evolution of a 
two-phase Fe-V system in a thermal gradient where the equilibrium concentration of both phases varies with temperature. 
In this case, the temperature range is $834^\circ C \to 1056^\circ C$.
The initial V compositions were 29.43 wt\% for the binodal region and 41.66 wt\% for the spinodal
region~\cite{dubiel2011sigma}.  The simulation details is provided in Section 3 of the supplementary document.

Figures~\ref{fig:darken_spino}(a-b)
display the corresponding microstructures for the binodal and spinodal compositions.
It consists of two phases: one is the solute-poor phase ($\alpha$), and another one is the solute-rich phase ($\sigma$). The equilibrium solute concentration of both $\alpha$ and $\sigma$ phases are given by the boundary of the miscibility gap in the phase diagram~\cite{dubiel2011sigma}.
We observe the formation of the V concentration gradient within each phase as the system has a temperature gradient. 
We take the 1D solute concentration profile along the black dotted line bounded by a green box in the binodal microstructure. 
This 1D composition profile is shown in Figures~\ref{fig:darken_spino}c. Here, the red and green lines represent the equilibrium V concentration for $\alpha$ and $\sigma$
phases (calculated using the Fe-V phase diagram). Here, due to the presence of a thermal gradient, the V concentration, with time, will try to maintain the equilibrium value at each temperature in both phases. As a result, it is visible from Figure~\ref{fig:darken_spino}c, that the V concentration in the $\alpha$ phase nearly follows the red line (equilibrium V concentration of $\alpha$ phase), while the V concentration within the $\sigma$ phase follows the green line (equilibrium V concentration of $\sigma$ phase).  
Here, the V concentration is slightly higher in both phases wrt. to the equilibrium concentration.
The circular shape of the $\sigma$ gives rise to the curvature effect, which is also responsible for the observed increase in equilibrium concentration~\cite{MUKHERJEE20093947}.

\begin{figure*}[ht]
\centering 
\includegraphics[width=\linewidth]{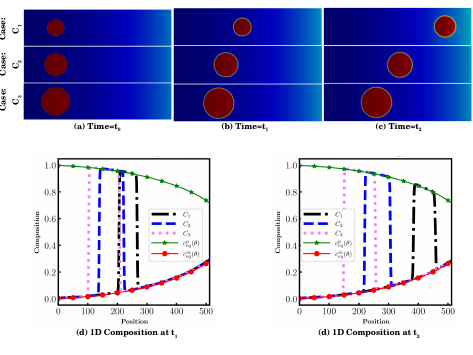}\caption{Microstructure evolution of single circular precipitate simulation for all three cases ($C_1$ to $C_3$ : top to bottom) at (a) $time=t_o$ (initial time), (b) $time=t_1$, and (c) $time=t_2$. The red region is the precipitate, and the remaining blue region is the matrix phase. In all these simulations, the right end has the highest temperature, while the left end has the lowest temperature.
1D composition plot taken at $y=\frac{ny}{2}$ for all three cases along with analytical ($c_{eq}^{p}(\theta)$) and ($c_{eq}^{m}(\theta)$) at (c) $time=t_1$ and (d) $time=t_2$. Here, $t_o<t_1<t_2$.}  
\label{fig:C_S_1}
\end{figure*}

The simulations with thermal gradient in a two-phase system show that 
a thermal gradient creates a concentration gradient inside each phase because of the temperature-dependent equilibrium concentration of phases. However, if the thermal gradient continues to exist even after establishing the concentration gradient, it raises another important question: how does the microstructure evolve in this scenario? 

 To address this query, we have selected a two-phase precipitate-matrix system for our simulations. The simulations are classified into three different categories:
\begin{enumerate}
  \item Single precipitate inside a matrix:
       \begin{enumerate}
           \item Precipitate with flat interface (Case: $R_1$, $R_2$ and $R_3$).
           \item Circular precipitate (Case: $C_1$, $C_2$ and $C_3$)
       \end{enumerate} 
  \item Two circular precipitates with different inter-precipitate distances (Case: $A$, $B$ and $C$).
          \item Multiple circular precipitates in a matrix.
\end{enumerate}

\begin{table}[ht]
\centering
\caption{Parameter details of the simulation model described in Section 2.}
\begin{tabular}[t]{lcc}
\hline
Parameters&Values\\
\hline
$A$  & 1.0\\
$M_c$  & 1.0\\
$\kappa_c$ & 1.0\\
$\theta_{low}$ & 0.1\\
$\theta_{high}$ & 0.9\\
$k_Q$ & 1.0\\
$\Delta X$ & 0.5\\
$\Delta Y$ & 0.5\\
\hline
\end{tabular}
\label{Tab:param_details}
\end{table}%

Starting here, our streamlined phase-field model (outlined in Section 2) is employed for efficient thermomigration analysis in a two-phase system. The parameter details are provided in Table~\ref{Tab:param_details}.
Note that in all cases, we consider the initial concentration to be the equilibrium solute concentration ($c_{p}^{eq}(\theta)$ and $c_{m}^{eq}(\theta)$ for the precipitate and the matrix phases, respectively ) as a function of temperature (spatially dependent) for each phase.
%If a point of the simulation domain has a temperature $\theta_p$ and the phase is precipitate, then the initial composition at this point is the equilibrium precipitate solute concentration at temperature $\theta_p$. 
\textcolor{black}{From here onward, the right-hand side of the simulation domain has the highest temperature ($\theta_{high}=0.9$), while the left end has the lowest temperature ($\theta_{low}=0.1$) as shown in Figure~\ref{fig:T_plot}.  }
We provide detailed explanations for each simulation in the following.

\begin{table*}
\centering
\caption{Simulation details for single, two and multi precipitate system.  }
  \begin{tabular}{lSSSSSS}
    \toprule
    \multirow{2}{*}{  Dataset} &
      \multicolumn{2}{c}{  Single Precipitate} & 
      \multicolumn{1}{c}{  Two Precipitates} &
      \multicolumn{1}{c}{  Multiple Precipitates} \\
      & {  Circular} & {  Flat interface} & {  Circular} & { Circular } \\
      \midrule
      Domain Size &  {$1024\Delta X \times 256\Delta Y$}  &  {$1024\Delta X \times 256\Delta Y$} &  {$1024\Delta X \times 256\Delta Y$} & {$1024\Delta X \times 1024\Delta Y$} \\
      $\theta$ &  {$0.1 - 0.9$} &  {$0.1 - 0.9$} &  {$0.1 - 0.9$}  &  {$0.1 - 0.9$}  \\
      Case &  {$R_1$, $R_2$,$R_3$} &  {$C_1$, $C_2$, $C_3$}  & {$A$, $B$, $C$}  & {-} \\
    \bottomrule
  \end{tabular}
  \label{Tab:sim_details_bc_t_case}
\end{table*}

\section{Results and Discussions}

\subsection{Single Precipitate}
We initially focus on a single precipitate model to eliminate the influence of inter-precipitate interactions. We examine two 
configurations: (a) an infinitely long precipitate in the $ Y$ direction having flat interface and (b) a circular precipitate. In both 
cases, we conduct three sets of simulations with varying sizes of precipitates. The simulation sets for the flat interface  
precipitate are denoted as $R_1$, $R_2$, and $R_3$, while for the circular precipitates, we label them as $C_1$, $C_2$, and 
$C_3$. The temperature range and total domain size remain consistent throughout all six cases, as 
specified in Table~\ref{Tab:sim_details_bc_t_case}. The only variation lies in the size of the precipitates, which increases 
from $R_1$ to $R_3$ and $C_1$ to $C_3$. 
Table~\ref{Tab:single_cases} contains a comprehensive summary of the initial positions and sizes of the single precipitate 
cases. For flat interface precipitates, the size corresponds to the width in the x-direction, while for circular precipitates, the size corresponds 
to the diameter of the precipitate.

\begin{table}[ht]
\centering
\caption{Details regarding the initial size and centre position of precipitate for both single flat interface and circular precipitate simulations.}
  \begin{tabular}{lSSSSSS}
    \toprule
    \multirow{1}{*}{  Case} 
      & {  Size} & {  Centre position} & {  Shape of precipitate} \\
      \midrule
      $R_1$ &  {$60$} &  {$(110,64)$} &  {flat interface}  \\
      $R_2$ &  {$80$} &  {$(110,64)$} &  {flat interface}   \\
      $R_3$ &  {$100$} &  {$(110,64)$} &  {flat interface}  \\
      $C_1$ &  {$60$} &  {$(110,64)$} &  {Circular}  \\
      $C_2$ &  {$80$} &  {$(110,64)$} &  {Circular}  \\
      $C_3$ &  {$100$} &  {$(110,64)$} &  {Circular}  \\
    \bottomrule
  \end{tabular}
  \label{Tab:single_cases}
\end{table}

\subsubsection{Flat interface precipitate}

We consider an initial flat interface precipitate embedded in a matrix to eliminate the effects associated with curvature~\cite{MUKHERJEE20093947}.  Figure~\ref{fig:R_S_1}(a) depicts the initial configuration of the flat interface precipitates for three cases, with $R_1$ positioned at the top and $R_3$ at the bottom.
The red region is the precipitate, while the blue colour region is the matrix phase.
The initial position of the precipitate is consistent across all cases.  Subsequently, Figures~\ref{fig:R_S_1}(b-c) depict the morphological changes observed at later time steps. As the system evolves, we observe the migration of the precipitate toward the region of higher temperature.
As mentioned earlier (in Section 3.1), as the temperature increases, the free energy in a system decreases, leading to the migration of solute atoms toward higher-temperature regions. 

 Analysis of the free energy versus composition plots at different temperatures (Figure~\ref{fig:model_f_c}) shows that a two-phase system exhibits lower free energy value at the higher temperature. Thus, in a two-phase system subjected to a thermal gradient, the state with the minimum free energy would involve the equilibrium of the two phases at the highest temperature point. In our case, the region with the highest temperature is on the right side. Hence, the precipitate initiates migration towards the higher temperature regions to achieve the configuration corresponding to the lowest free energy. During this migration, the precipitate-matrix interface must maintain the equilibrium solute concentration at each temperature, as illustrated by the phase diagram in Figure~\ref{fig:model_phase_diagram}. This equilibrium condition corresponds to the minimum free energy configuration at that particular temperature. However, the final minimum free energy configuration corresponds to the precipitate-matrix equilibrium at the highest temperature.

\begin{figure}[h!]
\centering 
\includegraphics[width=\linewidth]{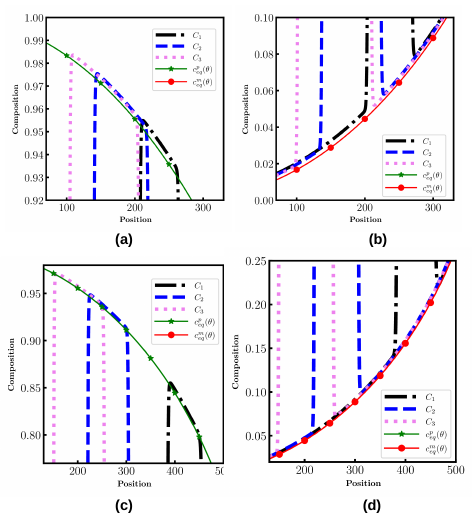}
\caption{Enlarge view of 1D compsoition plot for all three circular single precipitate along with analytical ($c_{eq}^{p}(\theta)$) and ($c_{eq}^{m}(\theta)$). (a) Showing the 1D composition inside precipitate along with ($c_{eq}^{p}(\theta)$) at $time=t_1$. (b) Showing the 1D composition at matrix adjacent to precipitate  along with ($c_{eq}^{m}(\theta)$) at $time=t_1$. (c) Showing the 1D composition inside precipitate along with ($c_{eq}^{p}(\theta)$) at $time=t_2$. (d) Showing the 1D composition at matrix adjacent to precipitate  along with ($c_{eq}^{m}(\theta)$) at $time=t_2$.   }
\label{fig:C_S_2}
\end{figure}

\begin{figure}[ht]
\centering 
\includegraphics[width=\linewidth]{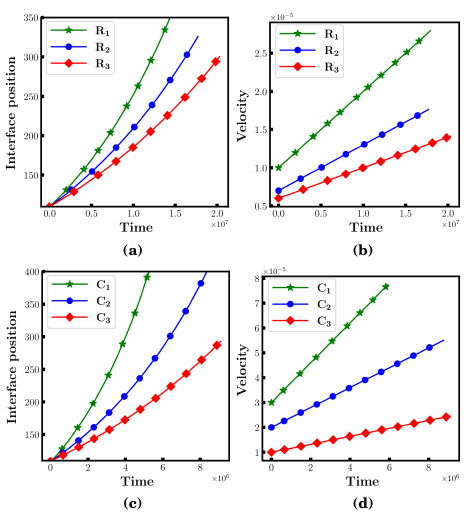}
\caption{ (a) Comparison of precipitate position with time for all three flat interface single precipitate simulations (Case: $R_1$, $R_2$, and $R_3$). (b) Comparison of migration velocity with time for all three flat interface single precipitate simulations (Case: $R_1$, $R_2$, and $R_3$). (c) Comparison of precipitate position with time for all three circular single precipitate simulations (Case: $C_1$, $C_2$, and $C_3$). (d) Comparison of migration velocity with time for all three circular single precipitate simulations (Case: $C_1$, $C_2$, and $C_3$). For both flat interface and circular precipitate, the velocity increases linearly with time.}
\label{fig:R_C_V_1}      % Give a unique 
\end{figure}

\begin{figure}[h!]
\centering 
   \includegraphics[width=0.8\linewidth]{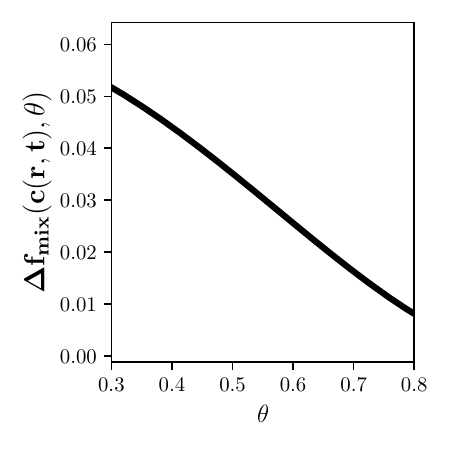}
   \caption{${\Delta}f_{mix}(c_{i}(\mathbf{r},t), \theta)$ vs $\theta$ at c=0.5, showing linear reduction in ${\Delta}f_{mix}(c_{i}(\mathbf{r},t), \theta)$ with increasing temperature ($\theta$).}
\label{fig:f_T_0.5_plot}      % Give a unique label
\end{figure}

\begin{figure*}[ht]
\centering 
\includegraphics[width=\linewidth]{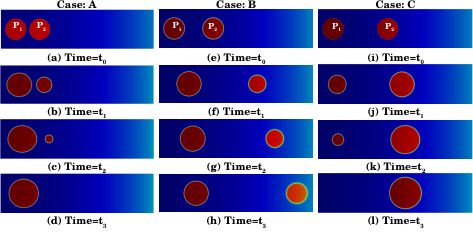}
\caption{ Microstructure evolution for two precipitates simulation for Case:A ($d_{inter}$ is small) at (a) $time=t_o$, (b) $time=t_1$, (c) $time=t_2$ and (d) $time=t_3$. In Case:A, $P_1$ grows and $P_2$ shrinks with time. 
Microstructure evolution for two precipitates simulation for Case:B ($d_{inter}$ is medium) at (e) $time=t_o$, (f) $time=t_1$, (g) $time=t_2$ and (h) $time=t_3$. In Case:B, initially $p_2$ shrinks but after some time its starts to grow. 
Microstructure evolution for two precipitates simulation for Case:C ($d_{inter}$ is large) at (i) $time=t_o$, (j) $time=t_1$, (k) $time=t_2$ and (l) $time=t_3$. In Case:C, $P_1$ shrinks and $P_2$ grows with time.}
\label{fig:C_T_1}
\end{figure*}

We compare the evolution of a one-dimensional (1D) composition profile, taken along the centre position of the vertical axis, with the equilibrium compositions of the matrix ($c_{eq}^{m}(\theta)$) and the precipitate ($c_{eq}^{p}(\theta)$). Remarkably, we observe a perfect agreement in the composition evolution between the precipitate and matrix phases during the migration process, aligning with the analytical equilibrium compositions.
Since here we consider the precipitate with the flat interface, the curvature effect is absent, resulting in a perfect equilibrium between solute concentration and temperature throughout the evolution. Figures~\ref{fig:R_S_1}(d-e) illustrate the evolution of 1D composition at times $t_1$ and $t_2$, along with the corresponding equilibrium compositions $c_{eq}^{m}(\theta)$ and $c_{eq}^{p}(\theta)$.

%\begin{table}[ht]
%\caption{Simulation parameters of the systems}
%\begin{tabular}{ |p{4cm}||p{2cm}|p{2cm}|p{2cm}|  }
% \hline
% \multicolumn{4}{|c|}{ {Simulation parameters (non-dimensional)}} \\
% \hline
%  {Parameters}&  {flat interface precipitate} &  {Circular precipitate} & {Multiple precipitate}\\
% \hline
%  {Simulation box size} &  {$512 \times 128$}   & {$512 \times 128$}  & {$512 \times 256$}\\
% \hline
%   {Systems}&  {$R_1$, $R_2$,$R_3$}   & {$C_1$, $C_2$, $C_3$}    & {\textcolor{cyan}{put the diameter}}\\
% \hline
%   {Temperature range ($T/T_{c}$, where ${T_{c} = 573 K}$)}&  {$0.1 - 0.9$}   & {$0.1 - 0.9$}    & {$0.1 - 0.9$}\\
% \hline
%\end{tabular}
%\label{Tab:sim_tab}
%\end{table}

%The precipitate migration can be explained by considering the solute concentration gradient within the precipitate. The solute concentration at the interface between the precipitate and matrix varies with temperature. Specifically, the interface at lower temperatures exhibits a higher solute concentration than at higher temperatures, resulting in the diffusion of the solute atoms through the precipitate from the lower temperature interface toward the higher temperature interface. 
%Thus, the solute atoms undergo diffusion through the precipitate, moving from the interface at a lower temperature towards the one at a higher temperature.

As the temperature increases, the equilibrium solute concentration of the precipitate phase ($c_{eq}^{p}(\theta)$) decreases, while for the matrix phase ($c_{eq}^{m}(\theta)$) it increases. Let's consider a precipitate initially at position $X_i$ with a corresponding temperature of ${\theta_i}$. During a small time interval ($\Delta t$), the precipitate migrates to a new position ($X_f$) with a corresponding higher temperature of ${\theta_f}$ (${\theta_f} > {\theta_i}$), where ($X_f-X_i$), (${\theta_f}  - {\theta_i}$), and $\Delta t$ are very small. At ${\theta}_f$, for the precipitate phase the equilibrium solute concentration ($c_p^{eq}({\theta}={\theta}_f)$) is lower than the equilibrium solute concentration ($c_p^{eq}({\theta}={\theta}_i)$) at ${\theta}_i$ because ${\theta}_i<{\theta}_f$. So, as the precipitate migrates from temperature $\theta_i$ to $\theta_f$, it must reject the excess solute ($c_p^{eq}({\theta}={\theta}_i)-c_p^{eq}({\theta}={\theta}_f)$) to maintain equilibrium at the new temperature ${\theta}_f$. This rejection of solute leads to a supersaturated state in the adjacent matrix. Consequently, this extra solute re-enters into the precipitate from the supersaturated matrix, facilitating the growth of the precipitate. 
Now, the precipitate and matrix phases have attained equilibrium concentration at this new elevated temperature, accompanied by a growth in the precipitate size. The precipitate is now ready to migrate to the next higher temperature stage. This sequence of events will repeat at each temperature during migration, and the size of the precipitate will increase during the migration.

The percentage change in size of the precipitate with respect to its initial size, as the migration occurs, can be calculated using Equation.~\ref{eq:size_chng}.
\begin{equation}
    \frac{d_f-d_i}{d_i}= \left\{\frac{(c_{eq}^{p}({\theta}={\theta}_{i}^{c})-c_{eq}^{m}({\theta}={\theta}_{i}^{c}))}{(c_{eq}^{p}({\theta}={\theta}_{f}^{c})-c_{eq}^{m}({\theta}={\theta}_{f}^{c}))}-1\right\} \times 100, 
\label{eq:size_chng}
\end{equation}
here $d_i$ denotes the initial precipitate size at the onset of thermomigration, $\theta_{i}^{C}$ signifies the central temperature point at the initial time, $d_f$ represents the final precipitate size following migration, and $\theta_{f}^{C}$ corresponds to the new central temperature point. The comprehensive derivation of Equation~\ref{eq:size_chng} is outlined in ~\ref{appendix2}.
Moreover, we compare the percentage size changes in all three scenarios ($R_1$, $R_2$, $R_3$) in our simulations with the analytical solution. A precise alignment is observed between the percentage size changes in the simulations and those derived from the analytical solution (Equation~\ref{eq:size_chng}), as depicted in Figure~\ref{fig:R_S_2}.
%The concurrence of the simulation results (\% change in size wrt. initial size) with the analytical calculation (using Eq~\ref{eq:size_chng}) provides validation for the aforementioned cyclic mechanism of precipitate migration driven by a thermal gradient.

%\textcolor{blue}{From Fig~\ref{one_rect_init0}-\ref{one_rect_init2} it is clearly visible that, for same amount of time $R_1$ migrated the furthest and $R_3$ shorted, while $R_3$ moderate distance from their initial position. Fig~\ref{fig:one_r_pos_vs_time} shows the center position with time. As we mentioned earlier the boundary conditions, temperature profile and mobility are same for three cases, so, it's evident that precipitate size influences  migration velocity. We calculated the velocity as precipitate migrates towards the higher temperature. Fig~\ref{fig:one_r_vel_vs_pos} shows the velocity vs position for all three cases. We can see that,  at any position, velocity for smaller precipitate is always higher compared to bigger precipitate.

%Fig~\ref{fig:one_rect_pot_1}-fig~\ref{fig:one_rect_pot_3} compare the chemical potential($df_{bulk}/dc$) among $R_1$, $R_2$ and $R_3$ at three different center positions(200, 250, and 300). There we found that, chemical potential is more for $R_1$ case and less steep for $R_3$ case. Since chemical potential is the driving force for diffusion, the size-dependent velocity is justified by the variations in chemical potential gradient among the three cases.%}
%(\textcolor{red}{There should be a linking between the earlier section and this section
%}  )   

\subsubsection{Circular Precipitate}

Figure~\ref{fig:C_S_1}(a) displays the initial setups of the circular precipitate-matrix systems, increasing in diameter from top to bottom ($C_1$ to $C_3$). The initial position of the precipitate is consistent across all cases.  Subsequently, Figures~\ref{fig:C_S_1}(b-c) depict the morphological changes observed at later time steps. As the system evolves, we observe the migration of the circular precipitate toward the region of higher temperature.
%To prevent precipitate dissolution, we maintain a slightly higher initial precipitate concentration (approximately 10\%) than the equilibrium precipitate composition ($c_{p}^{eq}\theta$).  Fig~\ref{fig:C_S_1}(b) through Fig~\ref{fig:C_S_1}(e) present the simulation outcomes at later time intervals.
We compare the evolution of the one-dimensional (1D) composition profile, taken along the center position of the vertical axis, with the equilibrium compositions of the matrix ($c_{eq}^{m}(\theta)$) and the precipitate ($c_{eq}^{p}(\theta)$) (shown in Figures~\ref{fig:C_S_1}(d-e)). Figures~\ref{fig:C_S_2}(a-d) represent the enlarged view of the 1D composition plot.  
In Figures~\ref{fig:C_S_2}(a-b), the composition plots depict the solute concentrations of the precipitate and matrix, respectively, at time $t_1$, while Figures~\ref{fig:C_S_2}(c-d) showcase the same at time $t_2$. At $t_1$ and $t_2$, the solute concentrations of both matrix and precipitate phases are slightly elevated compared to the equilibrium solute concentration ($c_{eq}^{p}(\theta)$ and $c_{eq}^{m}(\theta)$). 
The circular shape of the precipitate gives rise to the curvature effect, which is also responsible for the observed increase in equilibrium concentration for our simulation wrt. $c_{eq}^{p}(\theta)$ and $c_{eq}^{m}(\theta)$~\cite{MUKHERJEE20093947}.

\subsubsection{Velocity}

The observation of microstructural evolution in both flat interface (Figures~\ref{fig:R_S_1}(a-c)) and circular (Figures~\ref{fig:C_S_1}(a-c)) precipitates reveals that smaller precipitates migrated larger distance for same amount of time compared to larger ones. 
%We use constant and equal mobility in all these cases. Moreover, the temperature profile and boundary conditions remain consistent for all cases. Despite these similarities, the migration velocity is not the same for all sizes of precipitates.
The smaller precipitate ($R_1, C_1$) exhibits the fastest migration velocity, followed by moderate velocity for $R_2, C_2$, and the slowest velocity for $R_3, C_3$. 
Figure~\ref{fig:R_C_V_1}(a) illustrates the temporal evolution of the precipitate positions, while Figure~\ref{fig:R_C_V_1}(b) displays the velocity of the precipitate over time for all three single flat interface precipitates. Figures~\ref{fig:R_C_V_1}(c-d) display the exact quantities but for a circular precipitate. 
Notably, the migration velocity consistently increases over time at a constant rate as the precipitates migrate towards the higher-temperature region (shown in Figures~\ref{fig:R_C_V_1}(b and d)).
%As the temperature gradient remains constant for all three flat interface and circular precipitates, the driving force for migration is identical in all cases.
%However, it is evident from Figure~\ref{fig:R_C_V_1}(b) and Figure~\ref{fig:R_C_V_1}(c) that the migration rate is higher for smaller precipitates. 
The green line in these figures represents the velocity of the smaller precipitate, while the red line represents the velocity of the larger precipitate. As the precipitate migrates due to thermal gradient, it has to redistribute the excess solutes (resulting from the decrease in equilibrium solute concentration of precipitate with increasing temperature as explained in ~\ref{appendix2}) at each temperature to maintain equilibrium. The amount of total excess solute depends on the size of the precipitate. As a result, the smaller precipitate requires less time to achieve equilibrium at each temperature during migration by redistributing excess solutes (the total amount of excess solute for smaller precipitate is less than for the large precipitate). This disparity in solute redistribution time is the reason for the higher velocity observed in smaller precipitates than that in larger ones.

 Figure~\ref{fig:f_T_0.5_plot} shows the correlation between ${\Delta}f_{mix}(c_{i}(\mathbf{r},t), \theta)$ vs $\theta$ for a concentration of $0.5$. It is apparent that, in the given temperature range, the ${\Delta}f_{mix}(c_{i}(\mathbf{r},t), \theta)$ exhibits an almost linear decline with increasing temperature. In our specific scenario, a linear temperature gradient was applied throughout the entire domain, resulting in a linear proportional decrease in ${\Delta}f_{mix}(c_{i}(\mathbf{r},t), \theta)$ with distance. This linear reduction in ${\Delta}f_{mix}(c_{i}(\mathbf{r},t), \theta)$ acts as the propelling factor for the migration of the precipitate. As the linear reduction in ${\Delta}f_{mix}(c_{i}(\mathbf{r},t), \theta)$ remains consistent across temperature and distance, the precipitate experiences a steady acceleration, resulting in a uniform velocity increase as depicted in Figure~\ref{fig:R_C_V_1}(b) (for flat interface precipitate) and Figure~\ref{fig:R_C_V_1}(c) (for circular precipitate).

\subsection{Two precipitates}

In the preceding section, we obtained valuable insights into thermomigration phenomena in single precipitates of both flat interface and circular configurations. However, real microstructures consist of multiple precipitates of varying sizes, leading to Ostwald ripening~\cite{JUNG2023119167,HELL2023119095,FAN20021895}. So, it becomes essential to investigate the combined effects of Ostwald ripening and thermomigration on microstructure evolution. In the following sections, we present a comprehensive study that delves into the intricate interplay between coarsening and thermomigration of precipitates.

We initiate our study by examining a system composed of two identical circular precipitates with a diameter of 70 units embedded within a matrix. The system consists of two circular precipitates, namely $P_{1}$ (located on the colder side) and $P_{2}$ (located on the hotter side). We explore the impact on coarsening behaviour by varying the distance between the two precipitates ($d_{inter}$). Specifically, while keeping the position of $P_1$ fixed, we alter the center position of $P_2$, resulting in three distinct cases labelled as $A$, $B$, and $C$. Here, position refers to the center point of a precipitate. Table \ref{Tab:two_part_cc} presents detailed simulation parameters for each case.

%Two particles simulation is tested to understand the effect of thermomigration on coarsening behaviour.

\begin{table}[ht]
\centering
\caption{Two precipitate simulations details.}
  \begin{tabular}{lSSSSSS}
    \toprule
    \multirow{1}{*}{  Case} 
      & {Position of $P_1$} & {Position of $P_2$} & { $d_{inter}$} \\
      \midrule
      $A$ &  {$(50,64)$} &  {$(130,64)$} &  {$80$}  \\
      $B$ &  {$(50,64)$} &  {$(180,64)$} &  {$180$}   \\
      $C$ &  {$(50,64)$} &  {$(230,64)$} &  {$130$}  \\
    \bottomrule
  \end{tabular}
  \label{Tab:two_part_cc}
\end{table}

\subsubsection{Case:A}
In Case $A$, we examine a scenario where the inter-precipitate distance ($d_{inter}$) between $P_1$ and $P_2$ is extremely small. This configuration leads to the observation of shrinkage of $P_2$ and a slight migration towards the hotter side for both precipitates, as depicted in Figures \ref{fig:C_T_1}(a-d).
The presence of a thermal gradient induces migration in both precipitates. 
Nevertheless, the precipitate closer to the higher temperature terminal ($P_2$), undergoes faster migration than $P_1$.
 This temporal precedence occurs for a certain duration. As discussed in the previous section regarding single precipitate simulations, both precipitates release solute into the adjacent matrix during migration to maintain equilibrium concentration with the new temperature. However, in this Case, some of the solute rejected by $P_2$ in the inter-precipitate region is consumed by $P_1$ as it migrates to the higher temperature side. The transformation in the shape of $P_1$ is evident in Figures \ref{fig:C_T_1}(a-d). Consequently, the size of $P_1$ initially increases more than that of $P_2$. Subsequently, coarsening of $P_1$ and shrinking of $P_2$ occur due to the size differences, facilitated by the Gibbs-Thomson effect.

%We determined the deviation of the simulation composition from the equilibrium matrix composition, denoted as $C_m^{eq}(\theta)$, using Equation \ref{eqn:1_delc}. Figure \ref{fig:C_T_1}(m) illustrates the plot of $\Delta c$ (deviation in composition) against position for different time steps. In the region between the two precipitates, a solute $\Delta c$ gradient is observed, with higher values near the left interface of $P_2$ and lower values near the right interface of $P_1$. This gradient confirms that the $P_1$ precipitate consumes the solute rejected by $P_2$. Furthermore, it can be observed that as time progresses, the $\Delta c$ near the right interface of $P_2$ increases, while the $\Delta c$ near the left side of $P_1$ decreases. This phenomenon occurs due to the shrinkage of $P_2$ and the coarsening of $P_1$, leading to the manifestation of the Gibbs-Thomson effect.

%\textcolor{black}{
%\begin{equation}
%    \Delta c = C-C_{m}^{eq}(T),
%\label{eqn:1_delc}
%\end{equation}
%}

\subsubsection{Case:C}
In the case of $Case:C$, we consider a large inter-separation distance ($d_{inter}$) between $P_1$ and $P_2$. We observe a shrinkage in $P_1$ and coarsening of $P_2$, accompanied by the migration of $P_2$ towards the hotter side (shown in Figures~\ref{fig:C_T_1}(i-l)). The thermal gradient present in the system induces migration for both precipitates.

Throughout the migration process, each precipitate rejects excess solute into the surrounding matrix to establish equilibrium with the new temperature. In contrast to $Case:A$, most of the solute rejected by $P_2$ is absorbed by $P_2$ itself due to the substantial inter-precipitate distance. Two primary factors contribute to the larger size of $P_2$ compared to $P_1$. Firstly, the higher temperature of $P_2$ is crucial, as indicated by Equation~\ref{eqn:1314}, which illustrates the temperature-dependent equilibrium solubility of the precipitate $(c_{eq_i}^{p}(\theta))$. This equation reveals that at elevated temperatures, the change in equilibrium solute concentration wrt. temperature is more pronounced than at lower temperatures. Consequently, even for the same migration distance, the total solute rejected by $P_2$ precipitate surpasses that of $P_1$.

Additionally, due to its higher temperature location, $P_2$ precipitates migrate at a faster rate than $P_1$. Consequently, over the same duration, $P_2$ covers a greater distance, resulting in a more substantial temperature change and, consequently, increased excess solute rejection. The combination of these two factors leads to $P_2$ rejecting more excess solute than $P_1$.

Moreover, because of the considerable inter-precipitate distance, most of the solute rejected by $P_2$ re-enters the $P_2$ precipitate itself. This phenomenon causes a more pronounced increase in the size of the $P_2$ precipitate. Subsequently, the noticeable size disparity between the two precipitates leads to the larger precipitate ($P_2$) coarsening at the expense of the smaller precipitate ($P_1$).

%Fig~\ref{fig:C_T_1}(o) shows the evolution of $\Delta c$ with time for $Case C$.
%In between two precipitate regions $\Delta c$ is higher near the right 
%interface of $P_1$ and lower near the left interface of $P_2$. This $\Delta c$ gradient justify that 
%the solute diffusion from $P_1$ to $P_2$. Further, we can see that, as time goes on, 
%$\Delta c$ near the right interface of $P_1$ increase and left side of $P_2$ decreases.This is because 
%the $P_2$ shrink and $P_1$ coarsen causing the Gibbs-Thomson effect.  

\subsubsection{Case:B}
For $Case:B$, we consider the inter-separation distance between $P_1$ $\&$ $P_2$ to be moderate. 
Initially, we observe a shrinkage in $P_2$ and coarsening of $P_1$ along with its migration towards the hotter side (shown in Figures~\ref{fig:C_T_1}(e-g)) region, which is similar to $Case:A$. But as the system evolves, the velocity of $P_2$ increases because of both high temperature and smaller size compared to $P_1$. So, the inter-precipitate distance increases with time. After a certain point, the distance becomes sufficient for the rejected solute to re-enter into the $P_2$ itself (similar to Case:C). 
For higher temperatures, the equilibrium solute concentration change is more than lower temperatures for the same amount of temperature change. So, as  $P_2$ is on the higher temperature side, the change in equilibrium concentration with the temperature inside the precipitate is higher than $P_1$, which is on the lower temperature side. So, even for the same amount of migration, $P_2$ will grow more than $P_1$ as displayed in Figures~\ref{fig:C_T_1}(g-h). 
It is visible that, up to time=$t_2$, the size of $P_2$ decreases while the $P_1$ increases. At time=$t_2$ size of $P_2$, the precipitate is smaller than the $P_1$ precipitate. But after time=$t_2$, the size of the $P_2$ starts to grow. This growth of smaller precipitate (inverse coarsening) is opposite to the conventional Ostwald-ripening behaviour. 
The evolution of each precipitate size with time is provided in Section 4 of the supplementary material.

\begin{figure}[ht]
\centering 
\includegraphics[width=0.7\linewidth]{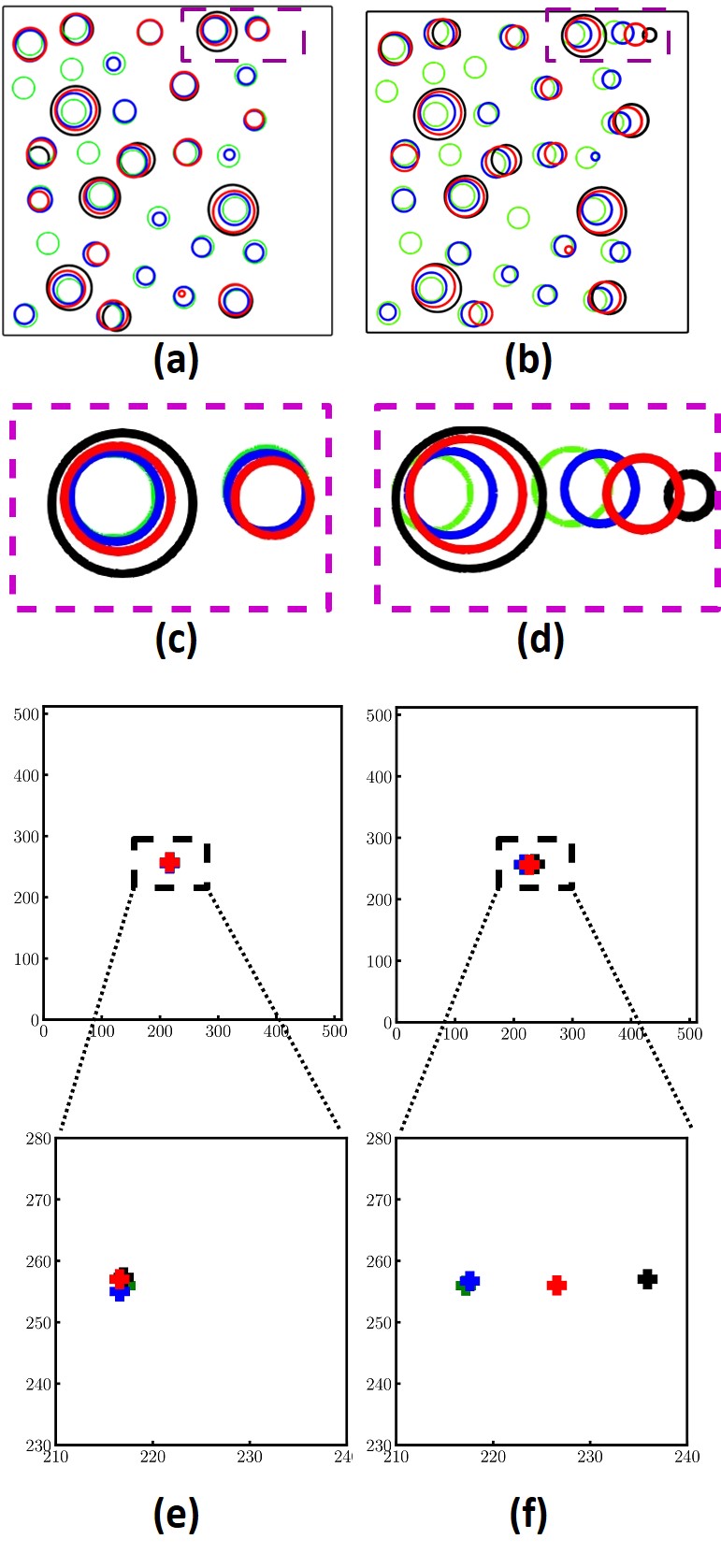}
\caption{Contour plot of precipitate matrix interface with time for (a) isothermal, (b) with thermal gradient at four different times ($t_o,t_1,t_2,\text{and }t_3$). Here, $t_o<t_1<t_2<t_3$.
Enlarged view of the flat interface
section for (c) isothermal case and (c) with thermal gradient case. 
Average centre($C_{avg}$) position of precipitates with time for (e) isothermal, (f) with thermal gradient. The plus signs of different colours represents the $C_{avg}$ for different times. Green colour to represent $time=t_o$, blue colour for $time=t_1$, red colour for $time=t_2$ and black colour for $time=t_3$.  }
\label{fig:C_M_1}
\end{figure}.

\subsection{Multiple precipitates}
In this section, we extend our model to investigate the behaviour of multiple circular precipitates within a matrix. We start our multiple precipitates simulations, carefully arranging their initial positions to prevent overlap. The sizes of the precipitates vary between 17 and 20 units, and they are positioned randomly within a $1024\Delta X \times 1024\Delta Y$ matrix. We perform two simulations, one under isothermal condition and another with an existing thermal gradient spanning 0.1 to 0.9. In the isothermal simulation, we set the temperature precisely at half the temperature range within the thermal gradient, specifically $\theta=0.5$.
The microstructural evolution is shown in Figure~\ref{fig:C_M_1}(a) for the isothermal case and Figure~\ref{fig:C_M_1}(b) for the thermal gradient case. 
Here, we plotted the contour lines representing the precipitate-matrix interface.
 The green contours denote the initial configuration at $t_o$, the blue contours represent the microstructure for at $t_1$, the red contours for time $t_2$, and the black contours for time $t_3$. Note that $t_0<t_1<t_2<t_3$.
 
In the case of isothermal conditions, shown in Figure \ref{fig:C_M_1}(a), we observe normal coarsening phenomenon over time. Larger precipitates coarsen, while smaller precipitates shrink.
However, in the presence of a thermal gradient, as the system evolves, the precipitates closer to the higher temperature region initiate migration due to the defined temperature gradient. In contrast, normal Gibbs-Thomson coarsening governs the behaviour of precipitates located in the lower temperature regime.
The microstructures during the later stages demonstrate the phenomena above, as shown in Figure \ref{fig:C_M_1}(b).
 As the precipitates in the high-temperature regions undergo migration, they encounter changes in size, as discussed in the section on two-precipitate simulations. The contour plots visually illustrate the simultaneous occurrence of precipitate coarsening and migration. We also observe that the velocity of precipitate migration is greater in the hotter regions than in the colder regions. Consequently, in regions with higher temperatures, thermomigration dominates, whereas in colder regions, Gibbs-Thomson coarsening takes precedence.
%For multiparticle simulation, a total number of 26 precipitates diameter in the range of 30 to 35 is set randomly inside a 512*512 matrix. 
%Here we found that precipitates on the higher temperature side migrate where, as the lower temperature side precipiate shows the normal Gibbs Thomson coarsening kinetics. Three different microstructure is shown at time $time=t_{0}$, $time=t_{1}$ and $time=t_{2}$ in fig a-c. The contour plot of the interfaces is shown in fig d. 

Figure~\ref{fig:C_M_1}(c-d) presents an enlarged view of the rectangular section indicated in Figure~\ref{fig:C_M_1}(a-b). It is evident that in the isothermal scenario (Figure~\ref{fig:C_M_1}(c)), the growth of the larger precipitate occurs at the expense of the smaller precipitate. Specifically, at time $t_4$, the smaller precipitate has completely dissolved, while the central positions of these precipitates remain nearly unchanged throughout the process.
In contrast, in the thermal gradient case (depicted in Figure~\ref{fig:C_M_1}(d)), the precipitate on the higher temperature side initiates migration. In this situation, both Ostwald ripening and thermomigration occur concurrently.

We conducted four sets of simulations with random precipitate distributions to determine the average center position ($C_{avg}$) for both the above-mentioned cases (isothermal and in the presence of thermal gradient).
These four initial microstructures are provided in Section 5 of the supplementary material.
We calculate the average center position of all precipitates over time for each set, and finally, we compute the average value of the center position by averaging the results from all four sets of simulations.
In the isothermal case, as shown in Figure \ref{fig:C_M_1}(e), $C_{avg}$ remains relatively stable over time, indicating minimal movement. However, in the presence of a thermal gradient, $C_{avg}$ shifts towards the higher temperature region, as shown in Figure \ref{fig:C_M_1}(f). This shift in $C_{avg}$ provides further evidence of the thermomigration of precipitates in addition to coarsening and shrinking.

\section{Conclusion}
\begin{itemize}

\item We developed a phase-field model to study the interplay between compositional and thermal gradients on microstructural evolution during thermomigration. The model is implemented in the MOOSE framework.

\item Simulated composition profiles 
within single-phase regions of Fe-N and 
Fe-C alloys show a remarkable match with those obtained by Darken and Oriani in their classical experiments on thermal diffusion in these alloys. Further, our simulated profiles compare well with the experimental results reported by Steiner in a Pb-Sn alloy where the thermal gradient spans both single and two-phase regions.

\item In two-phase systems, thermomigration produces translational motion of precipitates towards the high-temperature region. The migration rate inversely varies with the size of the precipitate.

\item The simulated composition profiles obtained from our study of thermomigration of a single precipitate show excellent agreement with the analytical solutions of our model.

\item By conducting systematic simulations of thermomigration using two precipitates within a matrix subjected to a given thermal gradient, we ascertain that the choices between growth and shrinkage of precipitate depend on the interparticle distance. Moreover, thermomigration can lead to inverse coarsening behavior leading to the growth of the smaller precipitate and the shrinkage of the larger one. It should be noted that such behaviour depends on the interparticle distance. This observation is highly significant as it contrasts with the isothermal Ostwald-ripening behavior of precipitates.

\item In multiple precipitate simulations, the presence of a thermal gradient induces thermomigration of precipitates in addition to capillarity-driven coarsening. 
When the thermal gradient is large, 
thermomigration dominates leading to the shift of the average center of mass of precipitates towards higher temperature regions.

In summary, our model accurately captures the coupled effects of thermomigration and diffusion-driven growth and coarsening of precipitates. %Thus our model can be used to gain crucial understanding of the thermomigration of micro-solder joints in advanced electronic packaging.

\end{itemize}

\section*{Acknowledgments}

The authors 
acknowledge National Supercomputing 
Mission (NSM) for providing computing 
resources of ``PARAM Sanganak'' at IIT 
Kanpur and ``PARAM Seva'' at IIT Hyderabad, implemented by C-DAC and 
supported by the Ministry of Electronics 
and Information Technology (MeitY) and 
Department of Science and Technology 
(DST), Government of India. 
Rajdip Mukherjee acknowledges financial support from
SERB core research grant (CRG/2019/006961).
S.B. acknowledges financial support from 
DST-NSM Grant DST/NSM/R\&D-HPC-Applications/2021/03. 
Soumya Bandyopadhyay thanks Indian Institute of Technology Kanpur for providing 
Institute Post Doctoral Fellowship.

\section*{Data Availability Statement}
The data that support the findings of this 
study are available from the corresponding 
authors upon reasonable request.

\section*{CRediT authorship contribution statement}
\textbf{Sandip Guin:} Conceptualization, Visualization, Methodology, Software, Investigation, Formal analysis, Validation, Data curation, Writing-Original Draft. \textbf{Soumya Bandyopadhyay:} Visualization, Methodology, Software, Investigation, Formal analysis, Validation, Data curation, Writing-Original Draft. \textbf{Saswata Bhattacharyya:} Supervision, Project administration, Resources, Writing - review $\&$ editing, Funding acquisition. $\&$ editing. \textbf{Rajdip Mukherjee:} Supervision, Project administration, Resources, Writing - review $\&$ editing, Funding acquisition.

\appendix
\section{Size change: Thermal migration}
\label{appendix2}

\begin{figure}[ht]
\centering 
\includegraphics[width=0.75\linewidth]{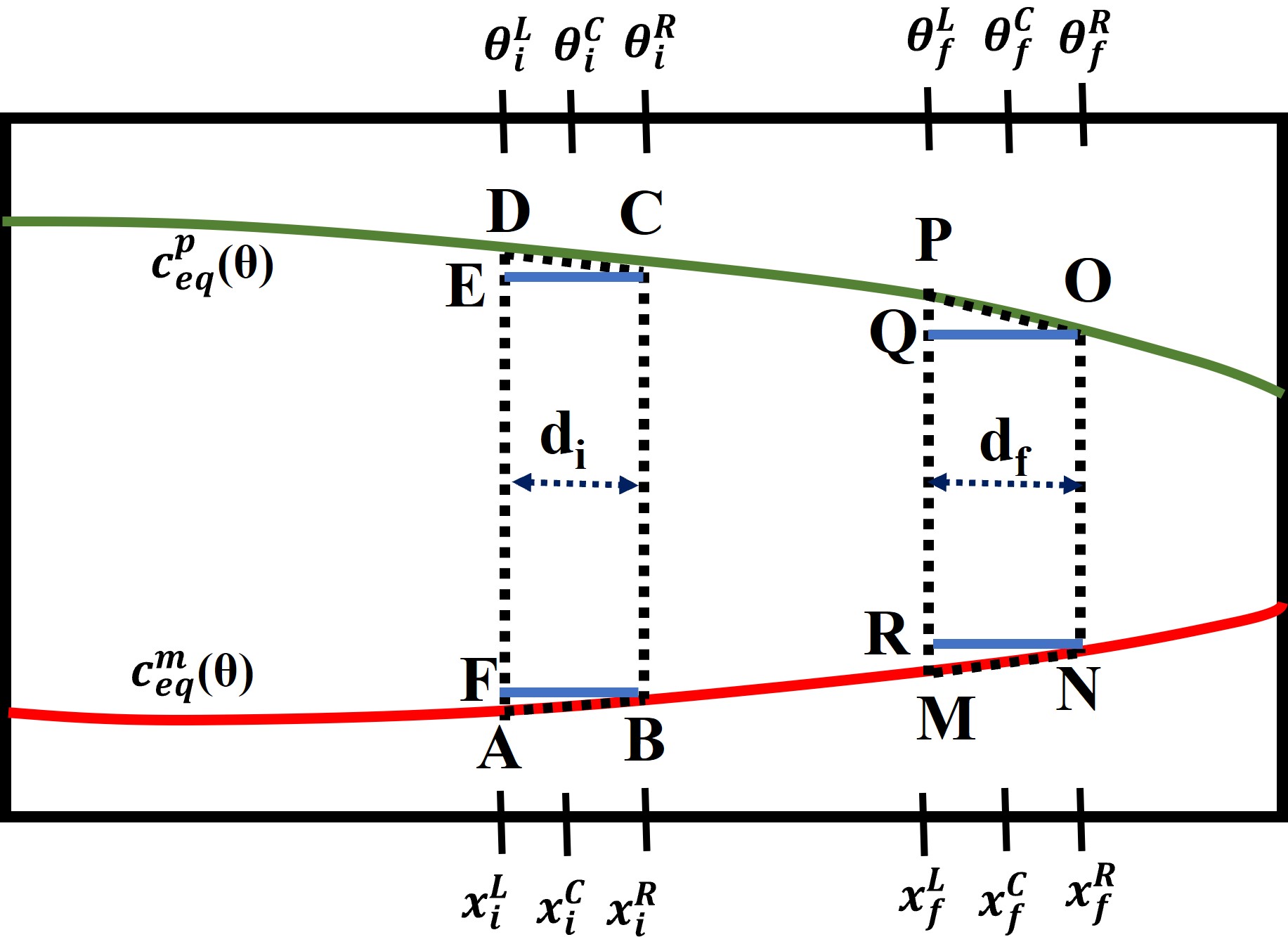}
\label{fig:size_chng_schematic}
\caption{Schematic of the composition profile in the matrix and precipitate
phases under themal gradient.}
\end{figure}

\textcolor{black}{Let consider a flat interface precipitate of size=$d_i$ inside a matrix under thermal gradient at time=$t_i$. Left interface of the precipitate is at $X_{i}^{L}$, right interface at $X_{i}^{R}$ and center line at $X_{i}^{C}$. This initial system is shown in Fig~\ref{fig:size_chng_schematic}. Now due to migration, the precipitate migrated to a new higher temperature at Time=$t_f$ and during the process size of the precipitate also changed to $d_f$. Now the Left interface of the precipitate is at $X_{f}^{L}$, right interface at $X_{f}^{R}$ and center line at $X_{f}^{C}$. Temperature corrosponding to point $X_{i}^{L}$, $X_{i}^{C}$, $X_{i}^{R}$, $X_{f}^{L}$, $X_{f}^{C}$ and  $X_{f}^{R}$  is ${\theta}_{i}^{L}$, ${\theta}_{i}^{C}$, ${\theta}_{i}^{R}$, ${\theta}_{f}^{L}$, ${\theta}_{f}^{C}$ and  ${\theta}_{f}^{R}$. We plot the analytical 1D composition plot in Fig~\ref{fig:size_chng_schematic}, where ABCD region is precipitate at time=$t_i$ and MNOP region is the precipitate at time=$t_f$.}

\textcolor{black}{If we consider AB and DC is linear At $time=t_i$, total area cover by ABCD region is given by:}

\begin{equation}
    A_{i} = d_i(AD+BC)
    \label{eq:B1}
\end{equation}

\begin{equation}
    AD =(c_{eq}^{p}({\theta}={\theta}_{i}^{L})-c_{eq}^{m}({\theta}={\theta}_{i}^{L}))
    \label{eq:B2}
\end{equation}

\begin{equation}
    BC =(c_{eq}^{p}({\theta}={\theta}_{i}^{R})-c_{eq}^{m}({\theta}={\theta}_{i}^{R}))
    \label{eq:B3}
\end{equation}

\textcolor{black}{replacing the values  of Eq~\ref{eq:B3} and Eq~\ref{eq:B3} in Eq~\ref{eq:B1}}

\begin{equation}
    A_{i} =d_i((c_{eq}^{p}({\theta}={\theta}_{i}^{L})-c_{eq}^{m}({\theta}={\theta}_{i}^{L}))+(c_{eq}^{p}({\theta}={\theta}_{i}^{R})-c_{eq}^{m}({\theta}={\theta}_{i}^{R})))
    \label{eq:B4}
\end{equation}

\begin{equation}
    A_{i} =d_i((c_{eq}^{p}({\theta}={\theta}_{i}^{L})+c_{eq}^{p}({\theta}={\theta}_{i}^{R}))-(c_{eq}^{m}({\theta}={\theta}_{i}^{L})+c_{eq}^{m}({\theta}={\theta}_{i}^{R})))
    \label{eq:B5}
\end{equation}

\begin{equation}
   (c_{eq}^{p}({\theta}={\theta}_{i}^{L})+c_{eq}^{p}({\theta}={\theta}_{i}^{R}))=2c_{eq}^{p}({\theta}={\theta}_{i}^{C})
    \label{eq:B6}
\end{equation}

\begin{equation}
   (c_{eq}^{m}({\theta}={\theta}_{i}^{L})+c_{eq}^{m}({\theta}={\theta}_{i}^{R}))=2c_{eq}^{m}({\theta}={\theta}_{i}^{C})
    \label{eq:B7}
\end{equation}

\begin{equation}
    A_{i} =2d_i(c_{eq}^{p}({\theta}={\theta}_{i}^{C})-c_{eq}^{m}({\theta}={\theta}_{i}^{C}))
    \label{eq:B8}
\end{equation}

\textcolor{black}{At $time=t_f$, total area cover by MNOP region is given by:}

\begin{equation}
    A_{f} =2d_f(c_{eq}^{p}({\theta}={\theta}_{f}^{C})-c_{eq}^{m}({\theta}={\theta}_{f}^{C}))
    \label{eq:B9}
\end{equation}

\textcolor{black}{As the total composition of the system is constant, from Eq~\ref{eq:B8} and Eq~\ref{eq:B9} we can write}

\begin{equation}
    A_{i} = A_{f}
    \label{eq:B10}
\end{equation}

\textcolor{black}{
\begin{equation}
    d_f = d_i\frac{(c_{eq}^{p}({\theta}={\theta}_{i}^{c})-c_{eq}^{m}({\theta}={\theta}_{i}^{c}))}{(c_{eq}^{p}({\theta}={\theta}_{f}^{c})-c_{eq}^{m}({\theta}={\theta}_{f}^{c}))}
\label{eq:B11}
\end{equation}
}

\begin{equation}
    \%\frac{d_f-d_i}{d_i}= \left\{\frac{(c_{eq}^{p}({\theta}={\theta}_{i}^{c})-c_{eq}^{m}({\theta}={\theta}_{i}^{c}))}{(c_{eq}^{p}({\theta}={\theta}_{f}^{c})-c_{eq}^{m}({\theta}={\theta}_{f}^{c}))}-1\right\}*100
\label{eq:B12}
\end{equation}

\bibliographystyle{unsrt}
\bibliography{scopus}
\end{document}

\begin{frontmatter}
\title{\textbf{Supplementary Material}\\Interplay between thermal and compositional gradients decides the microstructure during thermomigration: a phase-field study  
}
%Understanding Thermomigration through a Novel Phase-Field Model: From Single-Phase Evolution to Multi-Phase Dynamics in Solid-State Alloys

\author[a,b]{Sandip Guin$^\dagger$}
\author[a,c]{Soumya Bandyopadhyay$^\dagger$}
\author[d]{Saswata Bhattacharyya\corref{cor1}}
\ead{saswata@msme.iith.ac.in}
\author[a]{Rajdip Mukherjee\corref{cor1}}
\ead{rajdipm@iitk.ac.in}
\cortext[cor1]{Corresponding Authors}

\address[a]{Department of Materials Science and Engineering, Indian Institute of
Technology, Kanpur, Kanpur-208016, UP, India}

\address[b]{International College of Semiconductor Technology, National Yang Ming Chiao Tung University, Hsinchu 300, Taiwan} 

\address[c]{Department of Materials Science and Engineering, University of Florida, Gainesville, Florida-32611} 

%\address[c]{Department of Materials Science and Engineering, National Yang Ming Chiao Tung University, Hsinchu 300, Taiwan}
\address[d]{Department of Materials Science And Metallurgical Engineering
, Indian Institute of Technology, Hyderabad
, Sangareddy - 502285, Telangana, India}

\end{frontmatter}
\def\thefootnote{$\dagger$}\footnotetext{These authors contributed equally to this work}\def\thefootnote{\arabic{footnote}}

%% main text
\section{Single-phase system}
\label{sec:sample1}

Darken and Orani demonstrated the evolution of concentration gradient in single phase Fe-N and Fe-C alloys due to the presence of thermal gradient~\cite{DARKEN1954841}. 
To simulate their experimental results, we extracted the Gibbs free energy density data for the entire temperature range. For example, for the Fe-N sample, we have considered temperatures 
ranging from $600^oC$ to $756^oC$. For this temperature range, we have extracted the Gibbs free energy density vs composition data at different temperatures ranging from $600^oC$ to $756^oC$ using Calphad-based Thermocalc software~\cite{SUNDMAN1985153,MUNDHRA2024174288}. We have converted all these data into dimensionless values. 

\begin{figure}[ht]
\centering 
\includegraphics[width=0.99\linewidth]{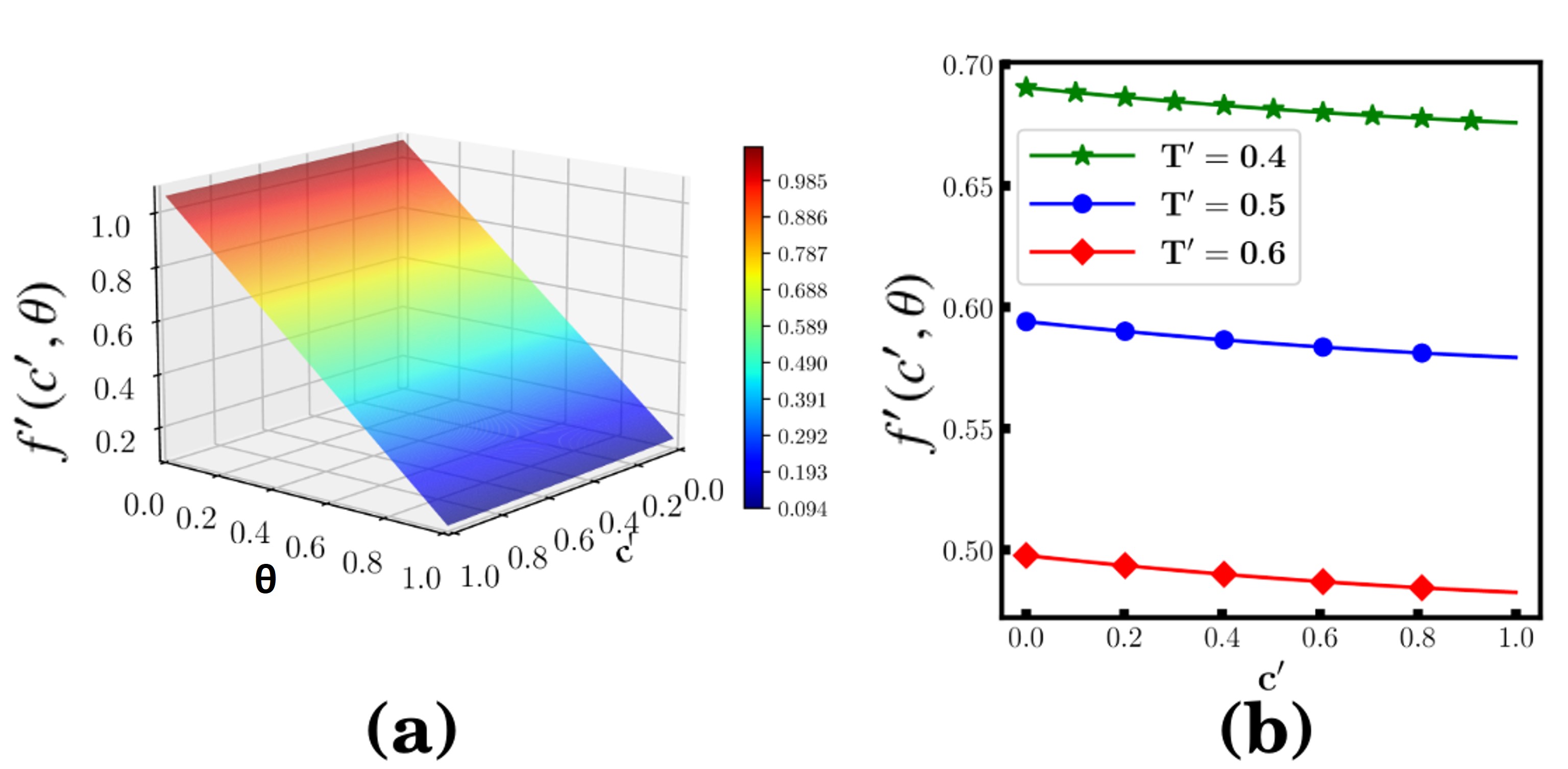}
\label{fig:f_scaled}
\caption{(a) $f'(c',\theta)$ plot as function of both $c'$ and $\theta$. (b) $f'(c',\theta)$ vs $c'$ plot at $\theta=$ 0.4, 0.5 and 0.6. }
\end{figure}

The composition is scaled between 0 and 1 using the following equation;

\begin{equation}
    c'= \frac{N(at\%)-c_{min}}{c_{max}-c_{min}},  
\label{eq:9}
\end{equation}
here, $c'$ is dimensionless form of concentration, $c_{min} = 0.002 at\%$ and $c_{max}= 0.5 at\%$.
The initial N concentration in the experimental case is 0.021 at\%~\cite{DARKEN1954841}.  
The temperature is scaled using the following equation; 

\begin{equation}
    \theta= \frac{T-T_{min}}{T_{max}-T_{min}},  
\label{eq:9}
\end{equation}
here,  $\theta$ is dimensionless form of temperature, $T_{min} = 600^oC$ and $T_{max}= 756^oC$.  For the experimental case, the temperature range is from $622^oC$ ($\theta=0.14$) to $756^oC$ ($\theta=1.0$). The free energy density is scaled using the following equation, 

\begin{equation}
    f'(c',\theta)= \frac{f(c,T)-f_{min}}{f_{max}-f_{min}},  
\end{equation}
here, $f'(c',\theta)$ is the scaled form of free energy density, $f_{min}=f(0.5 at\%, 756^oC)$ and $f_{max}=f(0.002 at\%,600^oC)$. Then, we have fitted $f'(c',\theta)$ into a function of $c'$ and $\theta$, which is given by;

\begin{equation}
    f'(c',\theta)= P{c'}^2+Qc'+R,   
\end{equation}
here, $P=-{0.0028\theta}^2+0.0046\theta+0.0049$, 
$Q=-0.0054\theta-0.0186$ and $R=-0.9637\theta+1.0759$. Figure~\ref{fig:f_scaled}(a) shows the $f'(c',\theta)$ surface plot as function of $c'$ and $\theta$. It is clearly visible that, with increasing $\theta$, $f'(c',\theta)$ value decreases. With increasing $c'$, $f'(c',\theta)$ also decreases, but this change is not clearly visible from the plot. For that reason, we have plotted a 1D profile of $f'(c',\theta)$ vs. $c'$ for $\theta=0.4,0.5,0.6$ (shown in Figure~\ref{fig:f_scaled}b). 
%In our main computational studies, we have used this $f'(c',\theta)$ equation in the case of $f(c_{i}(\mathbf{r},t), T)$ in Equation:15 in main article.

\begin{figure}[ht]
\centering 
\includegraphics[width=0.95\linewidth]{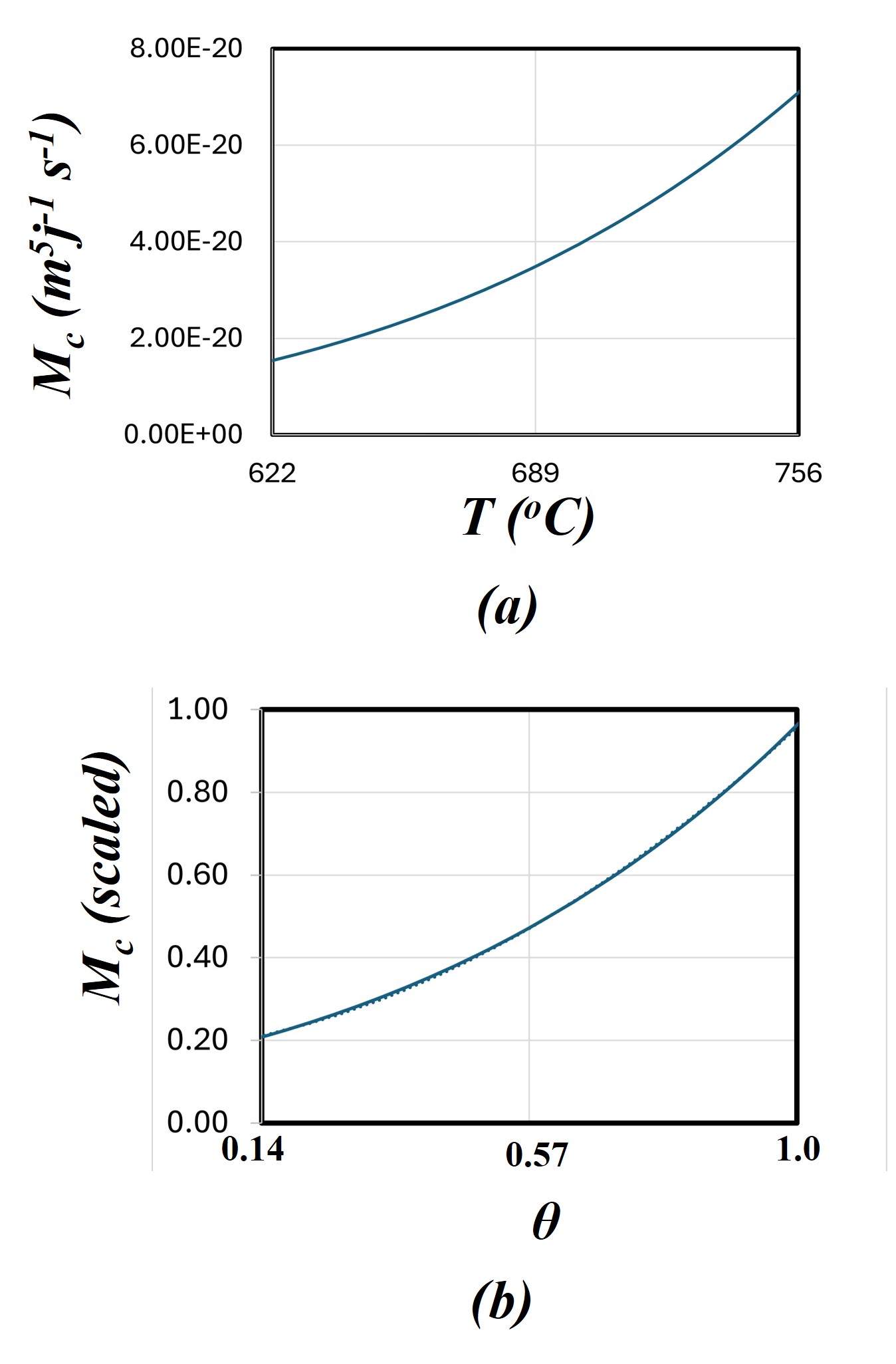}
\caption{(a) $M_c (m^5J^{-1}s^{-1})$ vs. $T (^oC)$ plot for N in $\alpha$-Fe. (b) Scaled mobility plot wrt. scaled temperature ($\theta$) for the same system. }
\label{fig:fen_mobility}
\end{figure}

We extracted the temperature-dependent mobility of N  in $\alpha$-Fe phase for Fe-N systems using the MOBHEA2 database in Thermocalc software~\cite{SUNDMAN1985153}. We have scaled these mobility values and used these scaled values as $M_c$ in Equation:15 in the main article.  Figure~\ref{fig:fen_mobility}(a) shows actual temperature-dependent mobility while, Figure~\ref{fig:fen_mobility}(b) shows the scaled mobility. It is visible from these figures that, the increment of mobility with temperature is similar for both actual and scaled mobility. In our simulation we, used the 
$M_c(scaled) = {0.611 \theta}^2+0.1714\theta+0.1749$.

For Fe-C, system we have done similar process as explained in the case for Fe-N system. We have extracted the Gibbs free energy density for the temperature range from $550^oc$ to $690^oc$.  Initial C concentration is 0.016at\%. We did the dimensionless conversion following the method as explained for Fe-N case. In this case $T_{min} = 550^oC$, $T_{max}= 690^oC$,  $c_{min} = 0.002 at\%$, $c_{max}= 0.5 at\%$, $f_{min}=f(0.5 at\%, 690^oC)$ and $f_{max}=f(0.002 at\%,550^oC)$.
Here also, we used scaled mobility of C in $\alpha$-Fe phase in our main simulation. 

%We use the temperature dependent diffusivity of N and C in $\alpha$-Fe phase which is given as:$D = D_0 \exp\left(-\frac{Q}{RT}\right)$, where $D_0$ is a constant, $Q$ is the activation energy, $R$ is the molar gas constant and $T$ is the temperature. For Fe-N system $D_0=2m^2/s$ and $Q=2KJ/mol$. For Fe-C system $D_0=2m^2/s$ and $Q=2KJ/mol$. 

For both cases, we use a simulation domain of $100\Delta X \times 40\Delta Y$, where initial concentration for N is at homogenous 0.021at\% and for C is 0.016 at\%. Constant temperature gradient is applied within the simulation domain, ranging from $622^oC$ to $756^oC$ for Fe-N system and $554^oC$ to $690^oC$ for Fe-C system. For both Fe-N and Fe-C system we take $\kappa_c$ to be 1.0.

\section{Pb-Sn system}

\begin{figure}[ht]
\centering 
\includegraphics[width=0.70\linewidth]{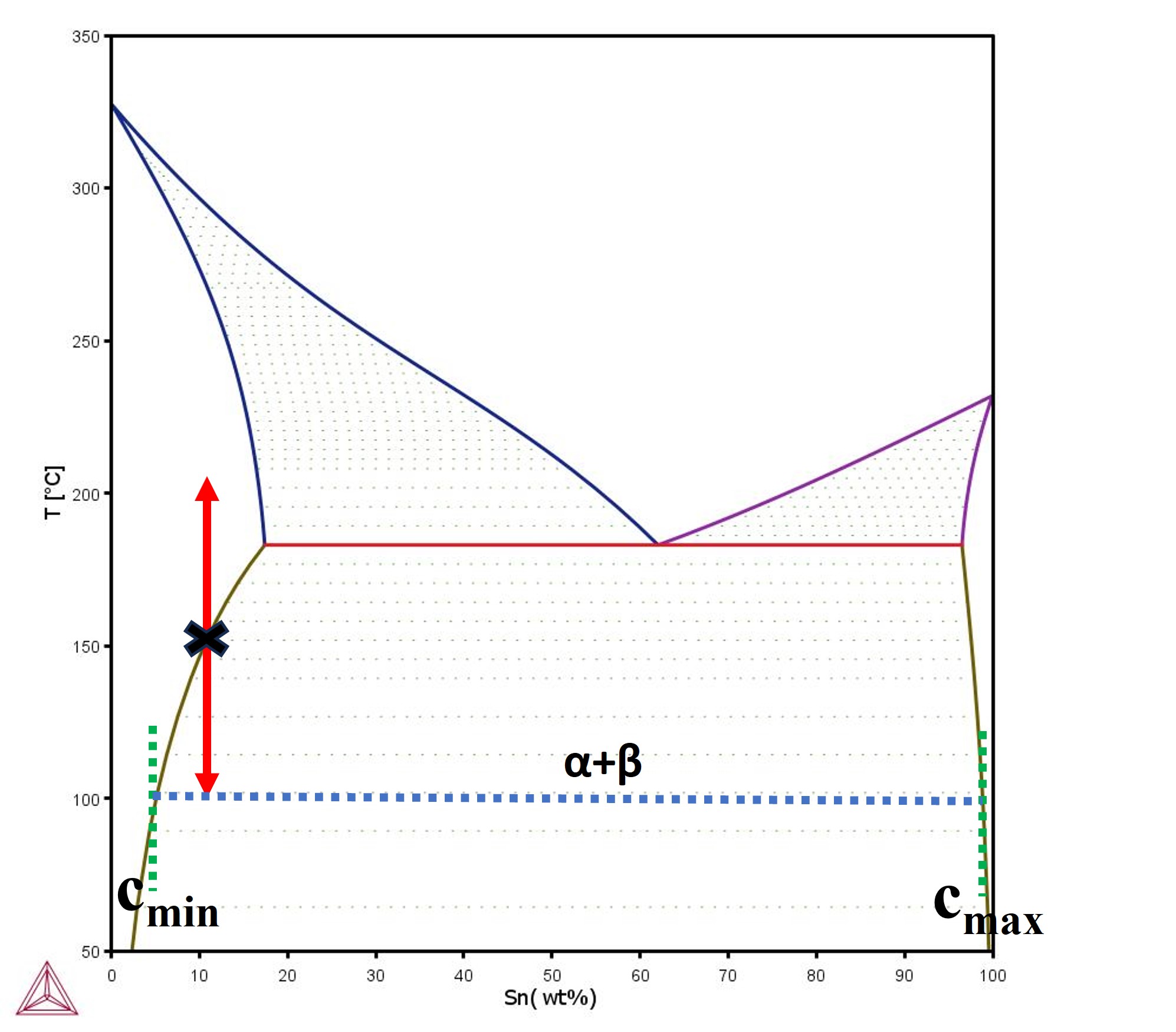}
\caption{Pb-Sn phase diagram~\cite{steiner19571}.}
\label{fig:pbsn}
\end{figure}

%For this case we choose Pb-Sn alloy as shown by Seiner et al. The phase diagram of Pb_Sn is shwn in . Blue dotted line is the initial alloy composition which is  Sn and rest is Pb. The temperature range is  $101^{\circ}C$ to $201^{\circ}C$. 

For the Pb-Sn system, the bulk free energy density is taken as; 
\begin{equation}
\begin{aligned}
  f(c,T) =  A(c_i-c_{eq_i}^{\alpha}
  (T))^2(c_i - c_{eq_i}^{\beta}(T))^2,
\label{eqn:12}
\end{aligned}
\end{equation}
here, $c_{eq_i}^{\alpha}(T)$ represents the equilibrium Sn concentration of the $\alpha$ phase as a function of 
temperature, while $c_{eq_i}^{\beta}(T)$ represents the equilibrium Sn concentration of the $\beta$ phase as a function of temperature. We have previously discussed the scaling of composition and temperature. 
Here, $T_{min}$ is equal to 101°C ($\theta=0$), and $T_{max}$ is equal to 210°C ($\theta=1$).
Therefore, $C_{min}$ corresponds to the equilibrium Sn concentration of the $\alpha$ phase at 101°C ($\theta=0$), and $C_{max}$ corresponds to the equilibrium Sn concentration of the $\beta$ phase at 101°C ($\theta=0$). $T_{min}$ is equal to 101°C, and $T_{max}$ is equal to 210°C. Pb-Sn phase diagram is shown in Figure~\ref{fig:pbsn}. Therefore, the final scaled free energy density can be expressed as follows;

\begin{equation}
  \begin{aligned}
  f'(c',\theta) = \\ & A(c'_i-(0.0181+0.1605\theta^2))^2 \\ & (c'_i - (0.9983-0.0146\theta^2))^2.
  \label{eqn:12}
  \end{aligned}
\end{equation}

\begin{figure}[ht]
\centering 
\includegraphics[width=0.80\linewidth]{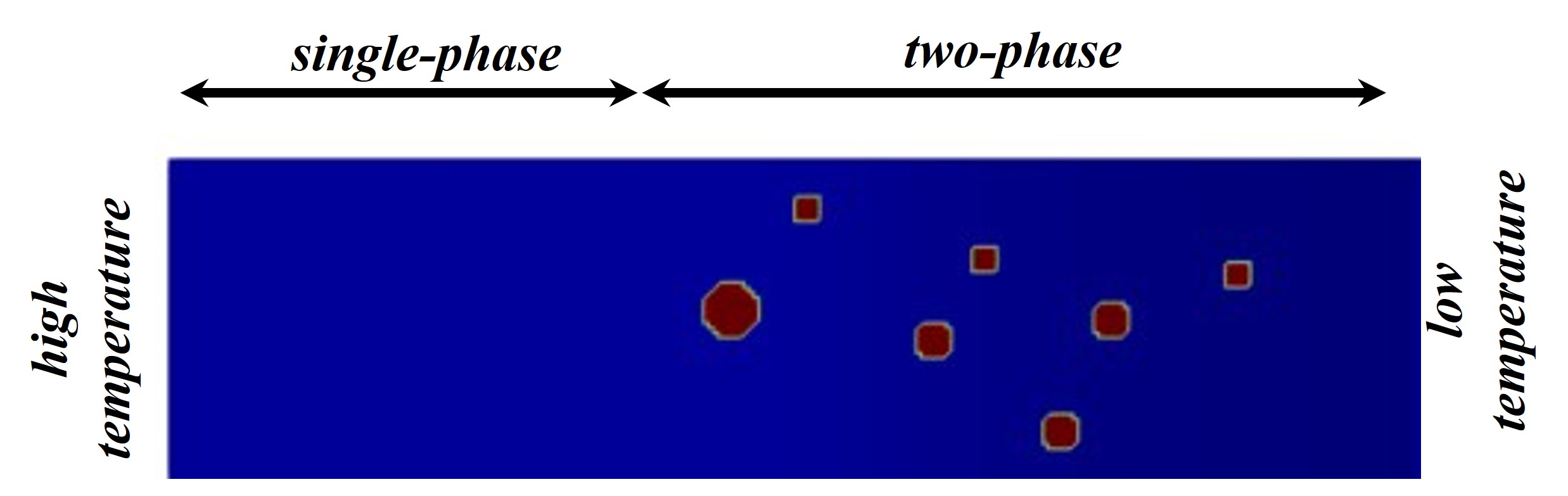}
\caption{Initial microstructure for Pb-Sn simulation}
\label{fig:pbsn_micro}
\end{figure}

We conducted four sets of simulations and obtained the average results for comparison with the experimental data.
At the initial time, the temperature is set to be the minimum temperature ($T=T_{min}$).
The initial microstructure consists of two domains: single phase-phase region and two-phase region. In single-phase region, the Sn concentration is kept $14 wt\%$, while in two-phase region we randomly placed $\beta$ precipitate with the $\alpha$ matrix. Here the volume fraction of the $\beta$ precipitate is according to the equilibrium volume fraction of $\beta$ precipiate at $T=T_{min}$ and 14 $wt\%$ Sn concentration. 
Figure~\ref{fig:pbsn_micro} shows the initial microstructure of one of simulations. 
Hence, the initial Sn concentration at this point was set to $14 wt\%$. After that, we apply the thermal gradient in the system.    

\begin{figure}[ht]
\centering 
\includegraphics[width=0.95\linewidth]{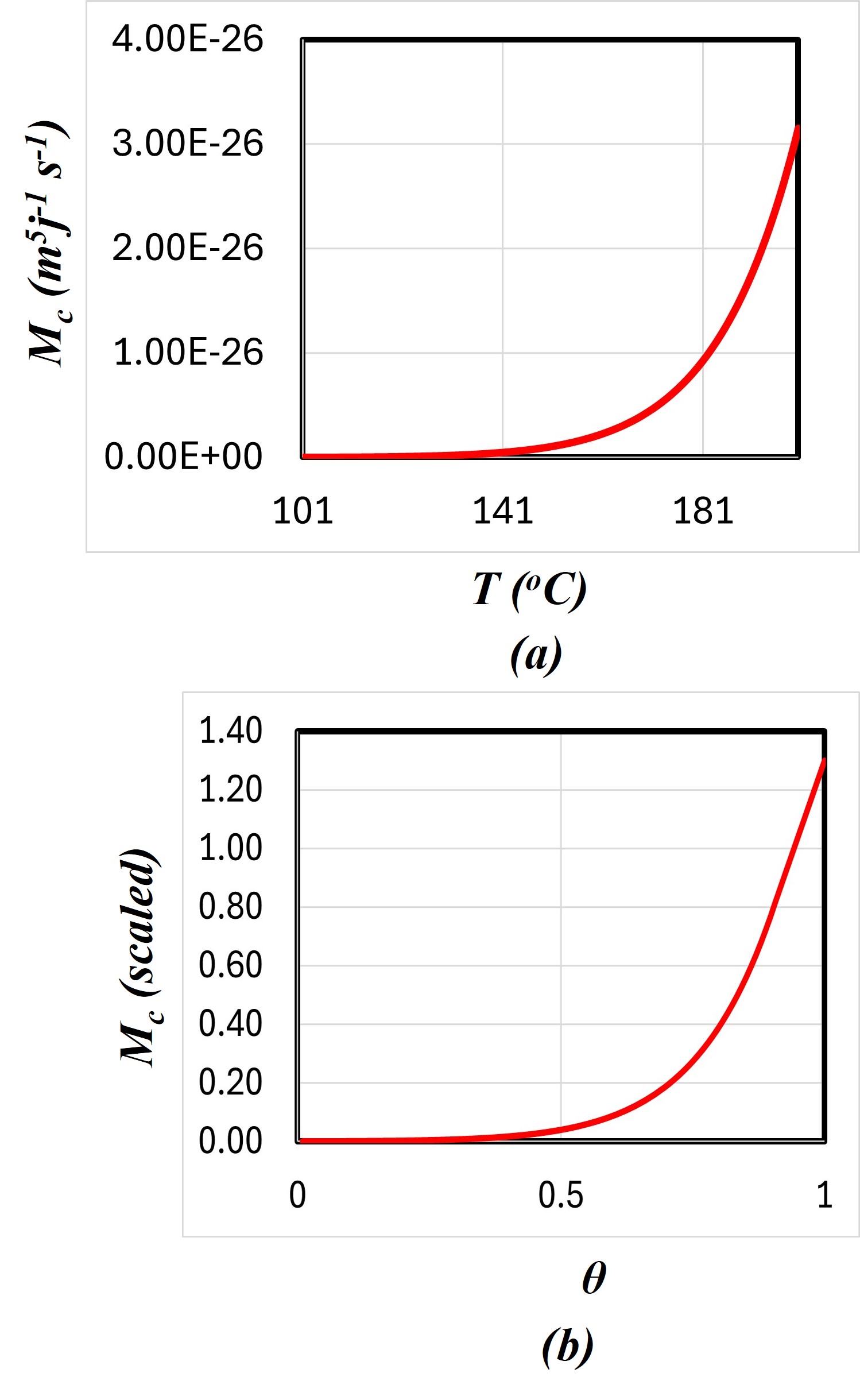}
\caption{(a) $M_c (m^5J^{-1}s^{-1})$ vs. $T (^oC)$ plot for Sn in Pb-Sn system. (b) Scaled mobility plot wrt. scaled temperature ($\theta$) for the same system. }
\label{fig:pbsn_mobility}
\end{figure}

In this case, we have also used the scaled mobility of Sn in Pb-Sn system. Figures~\ref{fig:pbsn_mobility}(a) shows the mobility of Sn in Pb-Sn system wrt. temperature, while Figures~\ref{fig:pbsn_mobility}(b) shows the scaled mobility for the same. In this case mobility of Sn in Pb-Sn system has been calculated using the MOBHEA2 database in Thermocalc software~\cite{SUNDMAN1985153}.
In our simulation, $M_c(scaled) = 4.02\theta^3-3.73\theta^2+1.009\theta-0.0557$. In this case our simulation size is For both cases, we use a simulation domain of $512\Delta X \times 128\Delta Y$. Here we consider $\kappa_c$ to be 1.0 for our simulation We run four simulation with different initial condition and take the average 1D composition plot which we have shown in the Figure 7b of main article. In all four simulations, only the initial position of the $\beta$ precipiates in the two-phase region are different. Figure~\ref{fig:pbsn_micro} shows one of the microstructure of such simulation.

\section{Fe-V system system}

\begin{figure}[ht]
\centering 
\includegraphics[width=0.80\linewidth]{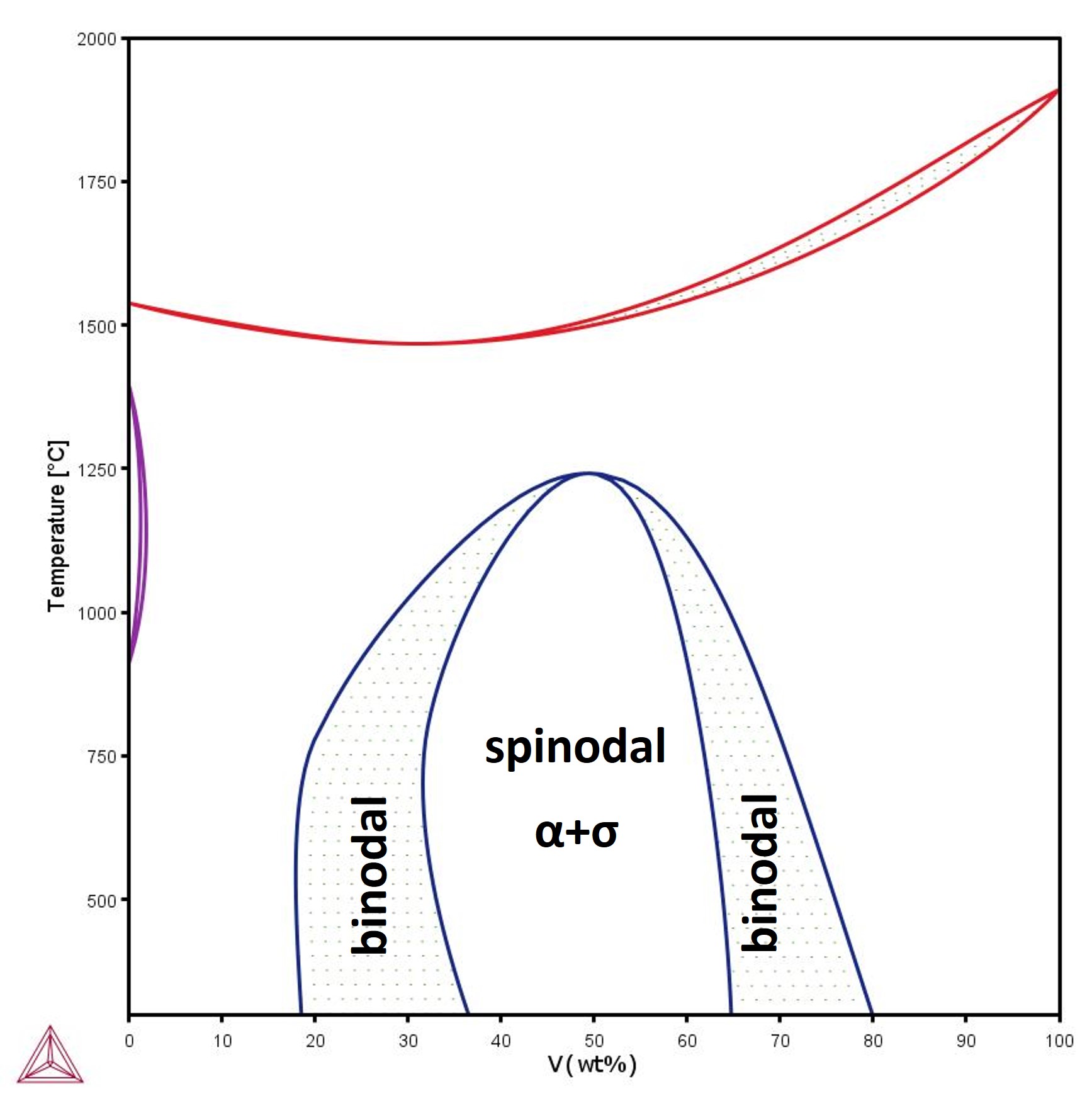}
\label{fig:pbsn}
\caption{Fe-V phase diagram}
\end{figure}

In this case, we followed a process similar to the Pb-Sn system, as discussed in the previous section (Section 3). 
However, in this case, $C_{min}$ corresponds to the equilibrium V concentration of the $\alpha$ phase at 834°C, and $C_{max}$ corresponds to the equilibrium V concentration of the $\sigma$ phase at 1056°C. Additionally, $T_{min}$ is equal to 834°C, and $T_{max}$ is equal to 1056°C. Therefore, the final scaled free energy density can be expressed as follows;

\begin{equation}
  \begin{aligned}
  f'(c',\theta)= \\ &  A(c'_i- (0.9959+0.0170\theta-1.4359\theta^2\\ &+2.5796\theta^3-1.5812\theta^4))^2 \\ &(c'_i- (0.0053+0.0292\theta+1.3499\theta^2 \\&-2.2324\theta^3+1.3674\theta^4))^2,
  \label{eqn:fe_v}
  \end{aligned}
\end{equation}
here, we take A=1.0. The value of $\kappa_{c}=1.0$ and $M_c=1.0$ in Equation:15 in the main article. The simulation domain size is $1024\Delta X \times 1024\Delta Y$.

\section{Case:B of two precipitate simulation}

\begin{figure}[ht]
\centering 
\includegraphics[width=0.7\linewidth]{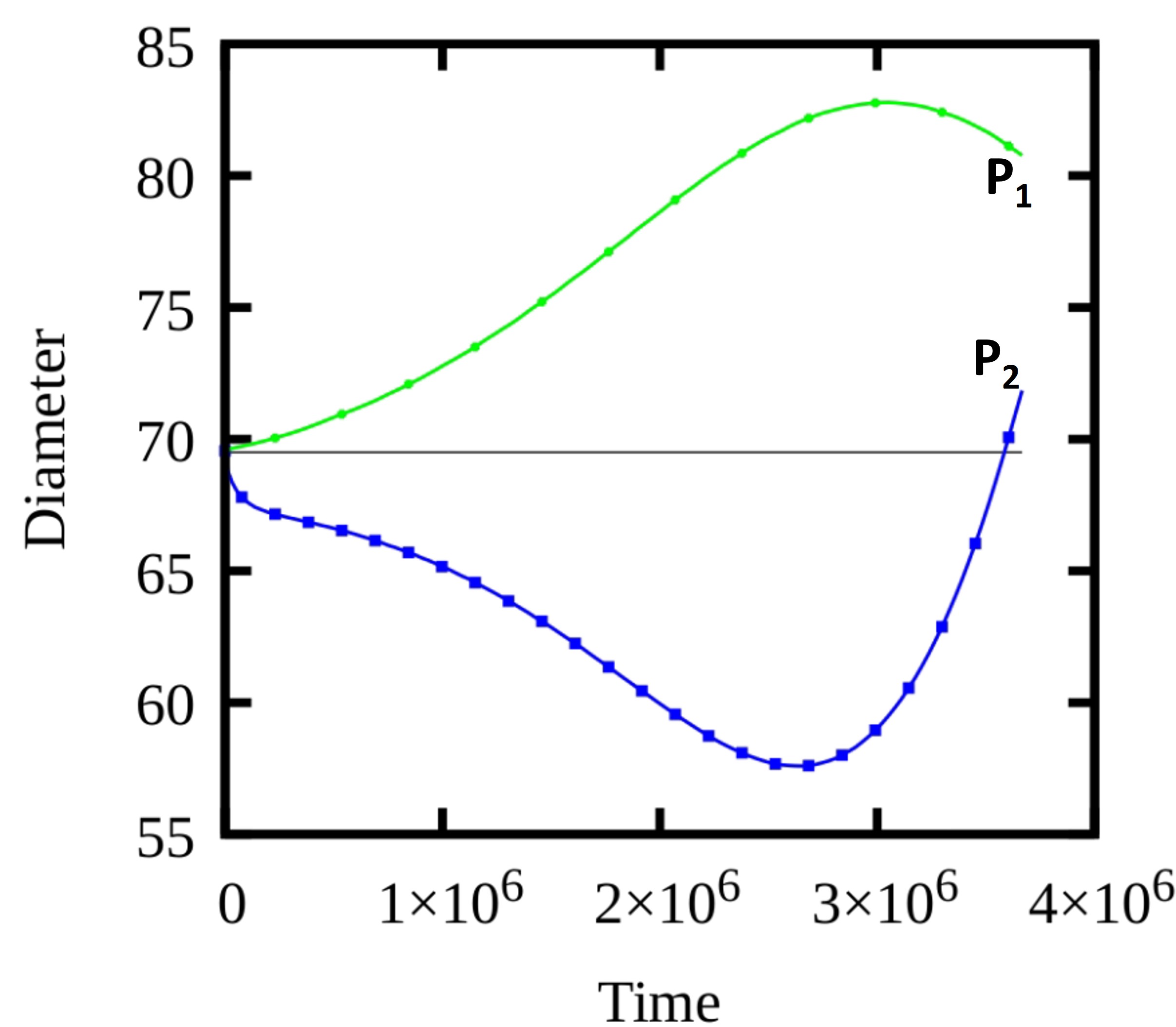}
\caption{Evolution of each precipitate size with time for Case:B of two precipitate simulations.}
\label{fig:two_ppt_dia}
\end{figure}

Figure~\ref{fig:two_ppt_dia} shows the evolution of each precipitate diameter with time for the Case:B of two precipitate simulations. In this case, $P_1$ and $P_2$ represent the left (lower temperature) and right (higher temperature) precipitate, respectively. 

\section{Multi-Precipitate simulations}

\begin{figure}[ht]
\centering 
\includegraphics[width=0.70\linewidth]{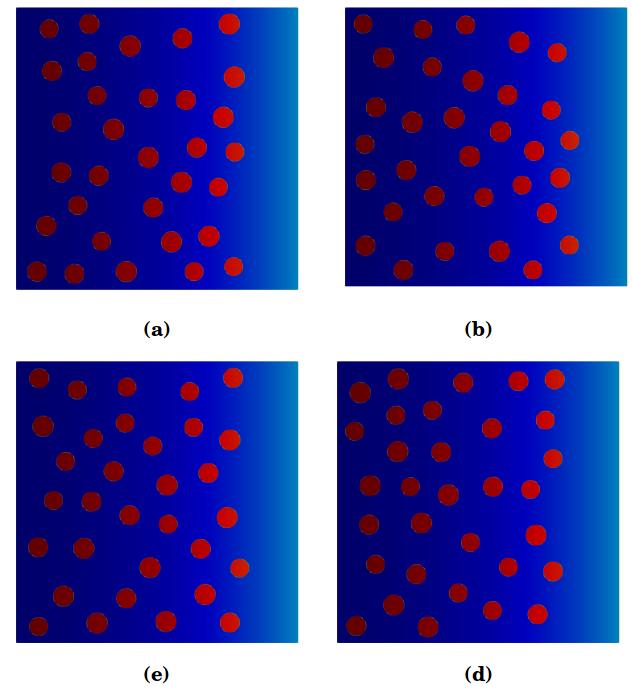}
\caption{Four sets of initial microstructure for multi-precipitates simulation.}
\label{fig:multi_ppt_4cases}
\end{figure}

In this case, we simulated four microstructures for both isothermal and thermal gradient cases. Precipitates are randomly placed inside the matrix. These initial microstructures are shown in Figure~\ref{fig:multi_ppt_4cases}. Here, the red circles represent the precipitates, and the blue region is the matrix. 

%% The Appendices part is started with the command \appendix;
%% appendix sections are then done as normal sections

\newpage
\bibliographystyle{unsrt}
\bibliography{scopus}

%% else use the following coding to input the bibitems directly in the
%% TeX file.

% \begin{thebibliography}{00}

% %% \bibitem{label}
% %% Text of bibliographic item

% \bibitem{}

% \end{thebibliography}